\documentclass[apj]{emulateapj}
\usepackage{apjfonts}
\usepackage{amsmath}
\usepackage{booktabs}
\begin{document}


%\title{The Relativistic Plasmoid Instability and Implication to Gamma Ray Pulse from Crab Pulsar}
\title{Evolution of the Relativistic Plasmoid-Chain in the Poynting-Dominated Plasma}

\author{Makoto Takamoto}
\affil{Max-Planck-Institut f\"ur Kernphysik, Heidelberg, Germany}
\email{makoto.takamoto@mpi-hd.mpg.de}

%\author{John, G. Kirk}
%\affil{Max-Planck-Institute fur Kernphysik, Heidelberg, Germany}
%\email{John.Kirk@mpi-hd.mpg.de}


%\altaffiltext{1}{Visiting Astronomer, Cerro Tololo Inter-American Observatory.
%CTIO is operated by AURA, Inc.\ under contract to the National Science
%\altaffiltext{2}{Society of Fellows, Harvard University.}
%\altaffiltext{3}{present address: Center for Astrophysics,
%    60 Garden Street, Cambridge, MA 02138}
%\altaffiltext{4}{Visiting Programmer, Space Telescope Science Institute}
%\altaffiltext{5}{Patron, Alonso's Bar and Grill}


\begin{abstract}
In this paper, 
%we investigate the evolution of the plasmoid-chain in high-$\sigma$ plasma. 
we investigate the evolution of the plasmoid-chain in a Poynting-dominated plasma. 
We model the relativistic current sheet with cold background plasma using the relativistic resistive magnetohydrodynamic approximation, 
and solve its temporal evolution numerically. 
We perform various calculations using different magnetization parameters of the background plasma 
and different Lundquist numbers. % with respect to the sheet length.  
Numerical results show that 
the initially induced plasmoid triggers a secondary tearing instability, 
%and the current sheet is gradually filled with plasmoids, that is, evolves into the plasmoid-chain 
which gradually fills the current sheet with plasmoids, 
as has also been observed in the non-relativistic case. 
%as predicted by non-relativistic work. 
%as non-relativistic work. 
We find 
%the plasmoid-chain highly enhances the reconnection rate 
%and the reconnection rate becomes independent of the Lundquist number 
%when the Lundquist number of the current sheet is larger than a critical value. 
the plasmoid-chain greatly enhances the reconnection rate, 
which becomes independent of the Lundquist number, 
when this exceeds a critical value. 
In addition, 
%we show the distribution of the plasmoid size becomes power law. 
we show the distribution of plasmoid size becomes a power law. 
Since magnetic reconnection is expected to play an important role 
in various high energy astrophysical phenomena, 
our results can be used for explaining the physical mechanism of them. 
\end{abstract}





\section{\label{sec:sec1}Introduction}

Magnetic reconnection is a process 
that converts magnetic field energy into thermal and kinetic energy very efficiently
~\citep{2000mrp..book.....B,2000mare.book.....P}. 
%Because of its high efficiency, 
Because of this, 
it is believed that 
magnetic reconnection plays an important role 
in various phenomena from the laboratory plasma to the astrophysical plasma. 
Recently, 
%the concern with properties of the relativistic magnetic reconnection has been growing, 
interest in the properties of relativistic magnetic reconnection has been growing, 
especially in Poynting-dominated plasmas, 
which are believed to be present in various high energy astrophysical phenomena, 
such as ultra relativistic jets \citep{2003ApJ...596L.159L,2011NewA...16...46B}, 
gamma ray bursts (GRB) \citep{2003astro.ph.12347L,2011ApJ...726...90Z}, 
and pulsar winds \citep{1984ApJ...283..694K,1984ApJ...283..710K,2001ApJ...547..437L,2003ApJ...591..366K}. 
In those models, 
%the Poynting-dominated plasma is assumed to be dissipated at some distance. 
the Poynting energy of the plasma is assumed to be dissipated into thermal and kinetic energy almost completely 
%at a distance. 
at some distance from the central object. 
However, 
such an efficient dissipation process is still unknown. 
%and many efficient dissipation processes are proposed
%There are alot of candidate of efficient dissipation processes
%Several studies have been made on finding efficient dissipation processes for this ten years 
In the last decade, 
several studies have been performed with the goal of finding efficient dissipation processes 
~\citep{1999ApJ...517..700L,2007MNRAS.380...51K,2012ApJ...755...76T,2012ApJ...760...43I,2013arXiv1303.2702A,2013arXiv1303.6434M,2013arXiv1303.6781M}. 
%Magnetic reconnection is one of the most probable models among them 
Magnetic reconnection is one of the most promising candidates among them, 
and has been studied actively from analytical~\citep{1994PhRvL..72..494B,2003ApJ...589..893L,2005MNRAS.358..113L} 
and numerical points of view
~\citep{2007MNRAS.374..415K,2009ApJ...696.1385Z,2009ApJ...705..907Z,2009JCoPh.228.6991D,2010ApJ...716L.214Z,2011ApJ...739L..53T,2012ApJ...750..129B}. 

For magnetic reconnection to be occurred, 
the plasma should contain current sheets. 
%The current sheet evolves into the Sweet-Parker sheet 
Such structures evolve into the Sweet-Parker configuration 
when the Lundquist number $S_L \equiv c_A L / \eta$ is small~\citep{2005PhRvL..95w5003L}, 
where $c_A$ is the Alfv\'en velocity, $L$ is the sheet length, $\eta$ is the resistivity. 
It is well-known that 
the reconnection rate of the Sweet-Parker sheet is very slow~\citep{1958IAUS....6..123S,1957JGR....62..509P,1963ApJS....8..177P}, 
so that 
%some other enhancement processes of the reconnection rate have actively investigated, 
a considerable number of studies have been conducted on finding an enhancement mechanism of the magnetic reconnection, 
such as the anomalous resistivity in the collisionless plasma~\citep{2005PhPl...12i2312U,2011PhPl...18k1206F}
and the turbulent effect~\citep{1999ApJ...517..700L,2009ApJ...700...63K}. 
Recently, 
it was found that 
%in current sheets with large Lundquist number, 
spontaneous current sheet fragmentation in a non-relativistic plasma occurs via secondary tearing instabilities 
%and the so called ``plasmoid-chain'' is formed 
when the Lundquist number exceeds a critical value, 
leading to the so-called plasmoid-chain.
The critical value is thought to be about $10^4$ in the non-relativistic plasma 
~\citep{2001EP&S...53..473S,2007PhPl...14j0703L,2009PhRvL.103j5004S,2010PhRvL.105w5002U,2011ApJ...737...24B,2012PhPl...19d2303L,2013arXiv1301.0331H,2013PhRvE..87a3102L}. 
In those works, 
it was shown that 
(1) the reconnection rate is enhanced by the plasmoid-chain and reaches typically $v_R \sim 10^{-2} c_A$; 
(2) the distribution of plasmoid size is either power law $w^{-p}$ or an exponential function $\exp [- w / \alpha']$ 
where $p$ is the power law index, $w$ is the plasmoid width, and $\alpha'$ is a constant. 
Since the plasma temperature in the plasmoid region is higher than that of background plasmas, 
the plasmoid-chain is expected to generate pulsed emissions. 
%Since plasmoids are usually hot comparing with the background plasma, 
%it can generate the pulsed profile of light curves. 
Hence, 
the plasmoid-chain is of interest from observational and theoretical points of view, 
especially in connection with the solar flare~\citep{2011ApJ...730...47B}. 

The first study of relativistic plasmoid-chain was given by \citet{2011MNRAS.418.1004Z}. 
They performed 2-dimensional and 3-dimensional numerical simulations of the relativistic magnetic reconnection 
using the relativistic resistive magnetohydrodynamic approximation \citep{2009JCoPh.228.6991D}. 
In those calculations, 
they assumed a background plasma 
with high Lundquist number $S_L \sim 10^5 - 10^8$, 
relativistic temperature $k_B T \sim m c^2$ and high magnetization parameter with respect to the mass density: 
$\sigma_m \equiv B_0^2 / 4 \pi \rho_0 \gamma_0^2 \sim 20$ 
where $B_0, \rho_0, \gamma_0$ are the background magnetic field in the laboratory frame, rest mass density and Lorentz factor, respectively. 
They also assumed the existence of a local anomalously large resistivity, 
so that their current sheet became very similar to the Petschek type one. 
They found that 
relativistic magnetic reconnection is similar to Petschek-type reconnection 
with a critical Lundquist number $\sim 10^8$, 
which is much larger than the non-relativistic cases. 
%They found that 
%the property of the magnetic reconnection is similar to the Petschek-type relativistic reconnection 
%and the critical Lundquist number is $\sim 10^8$, 
%which is much larger than the non-relativistic cases. 

In this paper, 
we investigate the evolution of the plasmoid-chain in a cold Poynting-dominated background plasma 
with large Lundquist number: $S_L \sim 10^3-10^5$. 
In particular, 
we mainly investigate statistical properties of the plasmoid-chain, 
such as the distribution function of the plasmoid width and the dynamics of X and O-points along the current sheet. 
To study the evolution of the secondary tearing instability, 
we use a uniform, constant resistivity, % with constant values, 
%and give perturbations to magnetic field only initially at the origin. 
and initialize the magnetic field with a perturbations localized at the origin. 
This enables us to understand the evolution of current sheets 
in which a tearing instability is triggered at a point. 
%However, 
%it is widely known that 
%it is very difficult to sufficiently dissipate the electromagnetic energy 
%by simple collisional Ohmic dissipation 
%within the observationally indicated characteristic times of phenomena, 
%so that many alternative mechanisms have been proposed~\citep{2003NewAR..47..667M,2006A&A...450..887G}. 
%Analogous problems can also be found in non-relativistic phenomena, 
%such as solar flares~\citep{1984SoPh...94..341S}, 
%and it seems to be a generic problem in high magnetic Reynolds number media. 
%Therefore, 
%research into mechanisms of efficient dissipation of electromagnetic energy is extensive. 

% magnetic field  = play important roles in high energy astrophysical phenomena

% reconnection

% fast reconnection = Petcheck, collisionless reconnection
%    <-> non relativistic work => plasmoid chain

%plasmoid chain  = as a origin of intermittent signals from solar flares

% difference from Zanotti & Dumbser 2011

% investigate the physics of plasmoid chain in Poynting dominated plasma

%In this paper, 
%%we use the following unit: $\mu = 1, c = 1$ where $\mu$ is the magnetic permeability and $c$ is the light velocity. 
%we use the following unit: $c = 1$ where $c$ is the light velocity. 

\section{\label{sec:sec3}Formation of Plasmoid-Chain}

In this section, 
we give a brief review of the non-relativistic plasmoid-chain theory. 

It is widely known that 
current sheets are unstable to the tearing instability. 
The maximum growth rate of this instability can be expressed as: 
$\omega_{max} = 1 / \sqrt{\tau_R \tau_A}$ 
where $\tau_R \equiv \delta^2 / \eta$ is the resistive diffusion timescale 
and $\tau_A \equiv \delta / c_A$ is the Alfven crossing time across a current sheet, 
$\delta$ is the sheet thickness, $\eta$ is the resistivity, 
and $c_A$ is the Alfv\'en velocity in the background plasma~\citep{1963PhFl....6..459F,1973ApJ...181..209L,2007MNRAS.374..415K}. 
This expression can be rewritten as follows: 
%\begin{eqnarray}
%  \label{eq:21}
%  \gamma_{max} &=& 1 / \sqrt{\tau_R \tau_A} = \left( \frac{\delta}{c_A} \frac{\delta^2}{\eta} \right)^{-1/2}
%%  = \frac{1}{\sqrt{c_A \delta / \eta}} \frac{c_A}{\delta} 
%  \nonumber
%  \\
%  &=& \frac{\tau_A^{-1}}{\sqrt{S_{\delta}}}
%  ,
%\end{eqnarray}
\begin{equation}
  \label{eq:21}
  \omega_{max} = 1 / \sqrt{\tau_R \tau_A} = \left( \frac{\delta}{c_A} \frac{\delta^2}{\eta} \right)^{-1/2} = \frac{\tau_A^{-1}}{\sqrt{S_{\delta}}}
  ,
\end{equation}
where $S_{\delta} = c_A \delta / \eta$ is the Lundquist number related to the sheet thickness $\delta$. 
This equation shows 
the tearing instability grows faster as the sheet thickness $\delta$ shrinks. 
The current sheet thickness behind plasmoids shrinks 
when the plasmoid grows along the current sheet, 
and this triggers the growth of other small plasmoids, 
which are called secondary plasmoids. 
Hence, 
we can expect that 
a current sheet would evolve into a stochastic plasmoid-chain 
in a few growth times of the largest plasmoid. 
%In \citep{2010PhRvL.105w5002U}, 
%Uzdensky et al (2010) assumed the existence of a critical Lundquist number $S_c$ 
\citet{2010PhRvL.105w5002U} assumed the existence of a critical Lundquist number $S_c$ 
at which current sheets become unstable to the plasmoid instability, 
%Here we use $S_L$ for the Lundquist number related to the sheet length $L$ 
%in order to distinguish from $S_{\delta}$ 
%that uses the sheet thickness $\delta$ for its characteristic scale length. 
and discussed the physical nature of the plasmoid-chain. 
This critical value introduces the smallest elementary structure in the chain, called the ``critical layer''; 
the related key parameters are the length scale $L_c = S_c \eta / c_A$, the thickness $\delta_c = L_c / \sqrt{S_c}$ and 
the reconnection rate $v_R = c_A / \sqrt{S_c}$. 
The authors also showed that 
the global reconnection rate is independent of the Lundquist number $S_L$ 
%when $S_L > S_c$ and the plasmoid-chain grows sufficiently in a current sheet. 
when $S_L > S_c$ and the plasmoid-chain reaches a statistical steady state. 
They obtained the global reconnection rate value as: $v_R \sim 10^{-2} c_A$ 
%by assuming the value of the critical Lundquist number as $S_c \sim 10^4$ 
%which is obtained by their numerical simulations. 
by assuming the value of the critical Lundquist number to be $S_c \sim 10^4$ 
in accordance with the results of their numerical simulations. 

%As is explained in \citep{2007PhPl...14j0703L,2009PhPl...16k2102B}, 
%As is explained by Loureiro et al (2007) and Bhattacharjee et al (2009)
As is explained by \citet{2007PhPl...14j0703L} and \citet{2009PhPl...16k2102B}, 
the above expression of the growth rate of the tearing instability $\omega_{max}$ can be reinterpreted as follows. 
When we consider the Sweet-Parker current sheet, 
we can obtain a relation between sheet thickness $\delta$ and sheet length $L$: 
$\delta \sim L / \sqrt{S_L}$ 
where $S_L \equiv L c_A / \eta$. 
Using this relation, 
the above equation can be rewritten as follows: 
\begin{equation}
  \label{eq:22_2}
  \omega_{max} \sim \frac{c_A}{L} \frac{L}{\delta} \sqrt{\frac{L}{\delta S_L}} \sim \frac{S_L^{1/4}}{\tau_{A,L}}
  ,
\end{equation}
where $\tau_{A,L} = L / c_A$. 
This equation means that 
the growth of the tearing instability becomes very fast 
when $S_L$ reaches about $10^4$, 
which they considered as the critical value of the Lundquist number. 
$S_L$ depends on the current sheet length $L$ 
and this means 
we need a very large numerical domain to study the effect of the plasmoid-chain. 

A more complete derivation is presented in \citep{2009PhPl...16k2102B,2010PhRvL.105w5002U,2013PhRvE..87a3102L}. 

\section{\label{sec:sec4}Numerical Setup}

\begin{table}[t]
  \center
  \begin{tabular}{ccccc} \toprule
    Name &  $\sigma_{in}$ & $c_A / c$  & $S_L \times 10^{-5}$ & $S_{\delta}$ \\
    \hline
    B1 & $0.14$  & $0.354$ & $1.13$  & $354$ \\
    B2 & $1.4$   & $0.767$ & $2.45$  & $767$ \\
    B3 & $14$    & $0.967$ & $3.09$  & $967$ \\
    B4 & $29$    & $0.983$ & $3.14$  & $983$ \\
%    H1 & $14$    & $0.967$ & $4 \times 10^{-3}$  & $0.73$  & $242$ \\
%    H2 & $14$    & $0.967$ & $16 \times 10^{-3}$ & $0.193$ & $60$ \\
    \hline
%  \begin{tabular}{lccccc} \toprule
%    Name &  $\sigma_{in}$ & $c_A / c$ & $\eta / c^2 t_0$ & $S_L \times 10^{-5}$ & $S$ \\
%    \hline
%    B1 & $0.14$  & $0.354$ & $10^{-3}$           & $1.13$  & $354$ \\
%    B2 & $1.4$   & $0.767$ & $10^{-3}$           & $2.45$  & $767$ \\
%    B3 & $14$    & $0.967$ & $10^{-3}$           & $3.09$  & $967$ \\
%    B4 & $29$    & $0.983$ & $10^{-3}$           & $3.14$  & $983$ \\
%%    H1 & $14$    & $0.967$ & $4 \times 10^{-3}$  & $0.73$  & $242$ \\
%%    H2 & $14$    & $0.967$ & $16 \times 10^{-3}$ & $0.193$ & $60$ \\
%    \hline
  \end{tabular}
  \caption{List of simulation parameters of basic runs. 
           $\sigma_{in} \equiv B_0^2 / 4 \pi \rho_0 h_0 \gamma_0^2$ is the magnetization parameter 
           where $B_0, \rho_0, h_0, \gamma_0$ are the upstream magnetic field, the rest mass density, the specific entharpy, and the Lorentz factor, respectively;  
           $c_A$ is the Alfv\'en velocity, 
%           $\eta$ is the resistivity, 
%           $t_0 \equiv \delta / c$ is the light crossing time of the sheet width, 
           $S_L \equiv L c_A / \eta$ is the Lundquist number related to the sheet length $L$ 
           and $S_{\delta} \equiv \delta c_A / \eta$ is the Lundquist number related to the sheet thickness $\delta$. }
  \label{list:1}
\end{table}

We model a very long current sheet using the relativistic resistive magnetohydrodynamic approximation. 
We use the resistive relativistic magnetohydrodynamics (RRMHD) scheme developed by \citet{2011ApJ...735..113T} 
extended to the multi-dimensional case using the unsplit method \citep{2005JCoPh.205..509G,2008JCoPh.227.4123G}. 
To preserve the divergence free constraint on the magnetic field, 
we use the constrained transport algorithm \citep{1988ApJ...332..659E}.  
%This scheme uses appropriate characteristic velocities for calculating numerical fluxes, 
%i.e., fast magnetosonic velocity and Alfv\'en velocity, 
%so that it describes turbulent flows accurately (see Takamoto \& Inoue 2011 for detail).
We calculate the RRMHD equations in a conservative fashion, 
and the mass density, momentum, and energy are also conserved within machine round-off error. 
For the equation of state, 
we assume a relativistic ideal gas with $h = 1 + (\Gamma / (\Gamma - 1))(p / \rho)$ 
where $\Gamma = 4 / 3$, 
$h$ is the specific relativistic enthalpy, 
$\rho$ is the rest mass density, and $p$ is the gas pressure. 
%Note that in this paper 
%we use the following unit: $\mu = 1, c = 1$ where $\mu$ is the magnetic permeability and $c$ is the light velocity. 
%we use the following unit: $c = 1$ where $c$ is the light velocity. 

\begin{figure*}[top]
 \centering
  \includegraphics[width=4.0cm,clip]{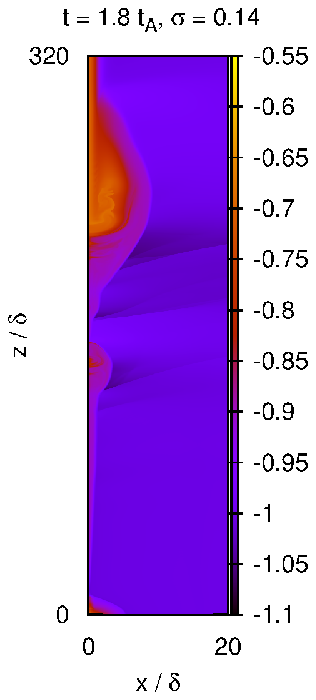}
  \includegraphics[width=4.0cm,clip]{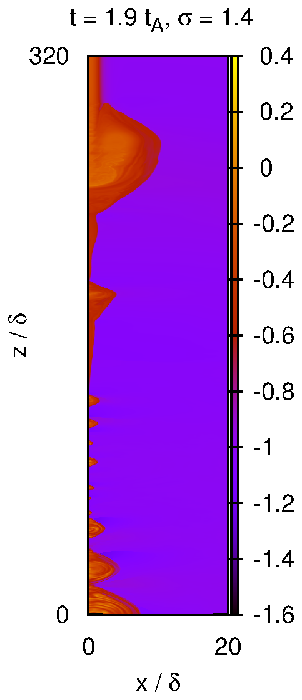}
  \includegraphics[width=4.0cm,clip]{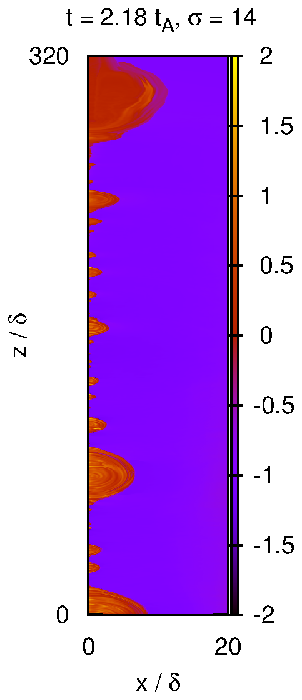}
  \includegraphics[width=4.0cm,clip]{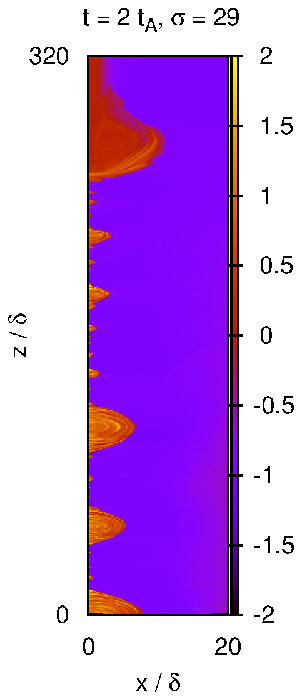}
  \caption{Snapshots of the temperature profile $\log_{10} [k_B T / m c^2]$ 
           of runs B1-B4 
           just before the largest plasmoid run away from the numerical domain 
           where $t_A = L_z / c_A$ is the Alfv\'en crossing time along the current sheet. 
          }
  \label{fig:5.1.1}
\end{figure*}

For our numerical calculations, 
%we prepared a square domain, $[L_x \times L_y]= [40 L \times 640 L]$ $[-L_x, L_x] \times [-L_y, L_y] = [-20 L, 20 L] \times [-320 L, 320 L]$ 
we prepare a square domain, $[0, L_x] \times [0, L_z] = [0, 20 \delta] \times [0, 320 \delta]$, 
where $\delta$ is the current sheet thickness. 
We divide it into homogeneous numerical meshes with size $\Delta = 5 \delta/128 \sim 0.04 \delta$ 
which is equivalent to the mesh number $N_x \times N_z = 512 \times 8192$. 
Note that to reduce computational costs 
we solve only a quarter region of the current sheet %the above numerical domain 
and impose the point symmetric boundary condition about $(x,z) = (0,0)$ following \citet{2009ApJ...705..907Z}. 
Hence, 
the above set up is equivalent to a square domain, $[-L_x, L_x] \times [-L_z, L_z] = [-20 \delta, 20 \delta] \times [-320 \delta, 320 \delta]$ 
divided by the mesh number $N_x \times N_z = 1024 \times 16384$
~\footnote{
Note that in Fig. \ref{fig:5.2.4} of Sec. \ref{sec:sec5.2}, 
we change the simulation box size $L_z$ to explore the property of the magnetic reconnection rate over a large parameter space of the Lundquist number. 
In the other part, 
we set $L_z = 320 \delta$. 
}. 
Along the boundaries $x = L_x$ and $z = L_z$, 
we impose the free boundary condition. 
For the initial condition, 
we consider the static relativistic Harris current sheet \citep{1966PhFl....9..277H,2003ApJ...591..366K}: 
\begin{eqnarray}
  B_z(x)  &=&  B_0 \tanh (x / \delta), \\
  p(x)    &=&  p_{in}    + p_s / \cosh^2 (x / \delta), \\
  \rho(x) &=&  \rho_{in} + \rho_s / \cosh^2 (x / \delta), 
  \label{eq:22}
\end{eqnarray}
where $p, \rho$ are the gas pressure and the rest mass density, 
and other variables are set to $0$ except for a small perturbation of the magnetic field described later. 
For the upstream region of the current sheet, 
we consider a cold plasma $\rho_{in} = 10 p_{in}$; 
for the inside of the sheet, 
we consider a relativistically hot plasma $\rho_s = p_s$ 
where $p_s = B_0^2 / 8 \pi$. 
Note that the temperature of the sheet decreases with decreasing magnetic field strength. 
%We determine the upstream gas pressure as $p_{in} = \beta p_B = \beta B_0^2 / 8 \pi$ 
%where $\beta$ is defined as the ratio of the gas pressure to the magnetic pressure. 
In this calculation, 
we use a constant resistivity $\eta$ to concentrate on investigating the effect of the plasmoid-chain on the reconnection rate. %for simplicity. 
Since it is easy to extend the law of resistivity, 
we will consider various kind of resistivity in our future work
\footnote{
In the case of a plasma with high temperature, 
the Coulomb collision cross section is usually very small 
and the collisional resistivity is also very small. 
However, 
if the plasma temperature rises up to the relativistic temperature $k_B T \sim m c^2$ 
where $k_B$ is the Boltzmann constant and $m$ is the particle rest mass, 
the photon density in the plasma becomes very dense 
and the Compton drag becomes effective as a dominant collisional process 
\citep{2008ApJ...688..555G}. 
}
. 
To trigger the initial tearing instability at the origin $(x,z) = (0,0)$, 
we add the following small perturbation to the magnetic field:
\begin{equation}
  \label{eq:23}
  \delta A_y = - 0.03 B_0 \delta \exp[-(x^2 + z^2) / 4 \delta^2]
  .
\end{equation}
Typical parameters used in our calculations are listed on Table \ref{list:1}. 
To model magnetic reconnection in high energy astrophysical phenomena, 
such as a relativistic jet, the Y-point of a pulsar magnetosphere and a gamma ray burst, 
we consider magnetically dominated plasma with magnetization parameter $\sigma_{in} > 1$. 
In the following sections, 
we present numerical results 
and consider the effects of the plasmoid-chain. % in the current sheet. 

\section{\label{sec:sec5}Results and Discussion}
In this section, 
we present numerical results of the tearing instability and evolution of the plasmoid-chain. 

\subsection{\label{sec:sec5.1}Temperature Profile}

Fig. \ref{fig:5.1.1} shows snapshots of temperature profiles of runs B1-B4 at the time 
when the largest plasmoid to result from the initial perturbation reaches the edge of numerical domain. 
%Since plasmoids move at the velocity approximately proportional to the Alfv\'en velocity of its upstream flow, 
Since plasmoids move at approximately the Alfv\'en speed of the upstream flow 
unless the plasmoid inertia is comparable to the magnetic field energy, 
%the escape time increases with decreasing its magnetization parameter $\sigma_{in}$. 
the escape time is of the order of $t_A$. 

First, 
we find that many plasmoids evolve along the current sheet. 
As we mentioned in the previous section, 
the evolution of a plasmoid induces a thinner current sheet behind it, 
%and it results in the secondary tearing instability and generates the plasmoids-chain. 
leading to a secondary tearing instability and the generation of a the plasmoids-chain. 
We also find that 
the thickness of the current sheet between plasmoids decreases 
and the apparent number of plasmoids increases 
with increasing magnetization parameter $\sigma$. 
We will discuss this in Sec. \ref{sec:sec5.2}. 
%This is because 
%the reconnection jet velocity becomes highly relativistic 
%when the upstream plasma $\beta$ is high as indicated in Appendix, 
%and the reconnection jet is directed in its flow direction 
%(Lyubarsky 2005, Zenitani et al ...). 
At the origin $(x,z) = (0,0)$, 
we note the existence of a large hot region. 
This is an artifact of our assumption of point symmetry about the origin, 
which means that 
%plasmoids moving downwards direction always have their counterparts moving in the opposite direction with the same magnitude of momentum; 
plasmoids entering the region from above have counterparts entering from below with the same magnitude and opposite speed. 
Their merger results in a plasmoid with zero momentum at the origin, 
which gradually accumulates matter as the simulation proceeds. 

%We note that 
%the apparent number of plasmoids increases with decreasing the plasma $\beta$. 
%In addition, 
%the largest plasmoid scale perpendicular to the current sheet seems to be independent of the plasma $\beta$; 
%on the other hand, 
%the scale parallel to the current sheet clearly decreases with decreasing the plasma $\beta$. 
%This can be understood using the Lorentz contraction. 
%As is explained in the following sections, 
%the Lorentz factor of plasmoid of run B1 is approximately $\gamma_p \sim 1.05$ 
%and that of run B4 is approximately $\gamma_p \sim 1.7$. 
%Using these values, 
%we expect the plasmoid length scale parallel to the current sheet of run B1 is $1.7$ times as large as that of run B4 
%if we assume the scale is the same in the plasmoid comoving frame. 
%Our numerical results agrees with this estimation. 

\subsection{\label{sec:sec5.2}Reconnection Rate}

\begin{figure}[h]
 \centering
  \includegraphics[width=7cm,clip]{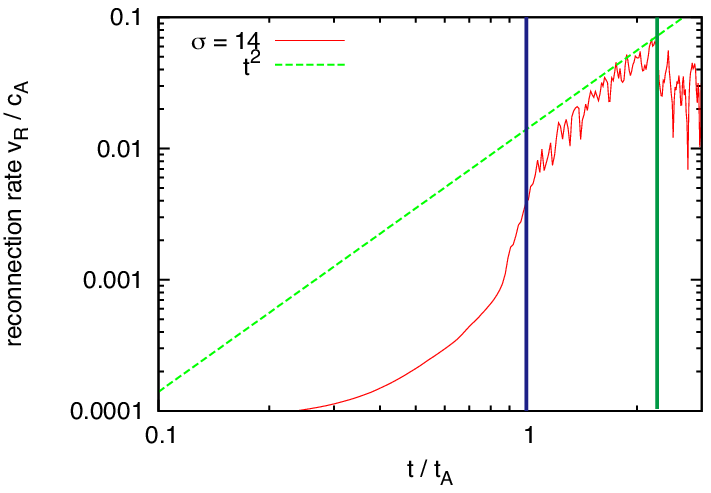}
  \includegraphics[width=7cm,clip]{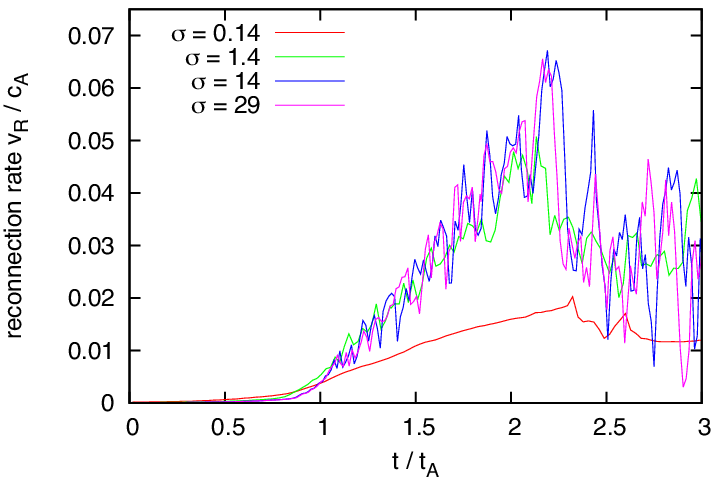}
  \caption{Top: The temporal evolution of the reconnection rate in the case of $\sigma_{in}= 14$. 
           The blue line at $t = t_A$ is the starting time of the plasmoid instability. 
           The green line at $t = 2.2 t_A$ is the time 
           when the largest plasmoid goes away from the numerical domain. 
           Bottom: The temporal evolution of the reconnection rate of runs B1-B4. 
          }
  \label{fig:5.2.1}
\end{figure}

Fig. \ref{fig:5.2.1} shows the time evolution of the reconnection rate 
in units of the Alfv\'en crossing time, $L_z / c_A \equiv t_A$. 
The reconnection rate is defined as: 
\begin{equation}
  \label{eq:24}
%  v_R \equiv - \frac{1}{L_z} \int^{L_z}_0 dz v_x(x = L_x, z) 
  v_R / c_A \equiv - \frac{c}{B_0 c_A L_z} \int^{L_z}_0 dz E_y(x = 0, z) 
  .
\end{equation}
The top panel is the result of B3 plotted using a logarithmic scale. 
%From the top panel, 
Here, 
we see that 
the evolution of the reconnection rate can be divided into three phases, 
separated in the figure by vertical lines at $t = t_A$ and $t = 2.2 t_A$. 
%We add 2 lines on the figure at $t = 350 \delta / c_A$ and $t = 700 \delta / c_A$. 
To the left of the blue line, 
the reconnection rate shows exponential growth due to the initial tearing instability at the origin. 
Between the blue and green lines, 
the reconnection rate oscillates around a power law growth rate with index approximately $2$. 
This is the region 
where the plasmoid-chain develops: %equivalent to the evolution of plasmoid-chain. 
%We find that 
%small plasmoids start to evolve at this stage 
%and this plasmoid-chain changes the evolution of the reconnection rate 
%from the exponential one to the power law due to its self similarity \citep{2010PhRvL.105w5002U}. 
small plasmoids start to appear, 
changing the growth rate from exponential to power law
\footnote{
This might be due to the self-similarity of structures in the plasmoid-chain~\citep{2001EP&S...53..473S,2010PhRvL.105w5002U}. 
}
. 
Finally, 
the reconnection rate saturates to the right of the green line, 
which marks the time when the largest plasmoid escape. 
%(see also the right figure of Figs. \ref{fig:5.2.3}). 
%It is very interesting problem to find the value of the reconnection rate due to the plasmoid instability 
%but, unfortunately, we cannot find it in this calculation. 

In the bottom panel, 
we compare reconnection rates of runs B1-B4. 
We find that 
runs B2-B4 show very similar evolution 
after the plasmoid instability is triggered. 
This indicates that 
the reconnection rate in units of the Alfv\'en velocity, $v_R / c_A$, becomes nearly independent of the magnetization parameter $\sigma$ 
in the strongly magnetized plasma, $\sigma_{in}> 1$, 
once the plasmoid instability starts. 
The reconnection rate grows until the largest plasmoid, 
which is initially triggered at the origin, 
escapes from the numerical domain, 
at which point the reconnection rate has increased up to $\sim 0.05 c_A$. 
%after the escaping of the plasmoid, 
After this, 
the plasmoid-chain reaches a statistical steady state 
and the averaged reconnection rate is about $0.03 c_A$, 
which is approximately twice that of the relativistic tearing instability without a plasmoid-chain~\citep{2011ApJ...739L..53T}. 
%We will discuss this value later. 
Note that 
the reconnection rate of run B1 is lower than that of other runs. 
This is because in this case 
the plasmoid-chain does not grow sufficiently as can be seen in the left panel of Fig. \ref{fig:5.1.1} 
where only one secondary plasmoid is generated. 
This reduces its reconnection rate comparing with other runs including the fully evolved plasmoid-chain. 
We discuss later the reason why the plasmoid instability does not grow in this case. 

\begin{figure}[h]
 \centering
  \includegraphics[width=6cm,clip]{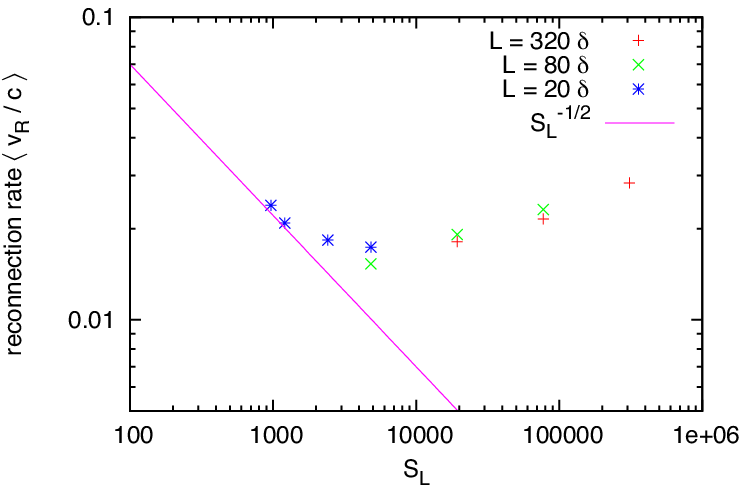}
  \includegraphics[width=6cm,clip]{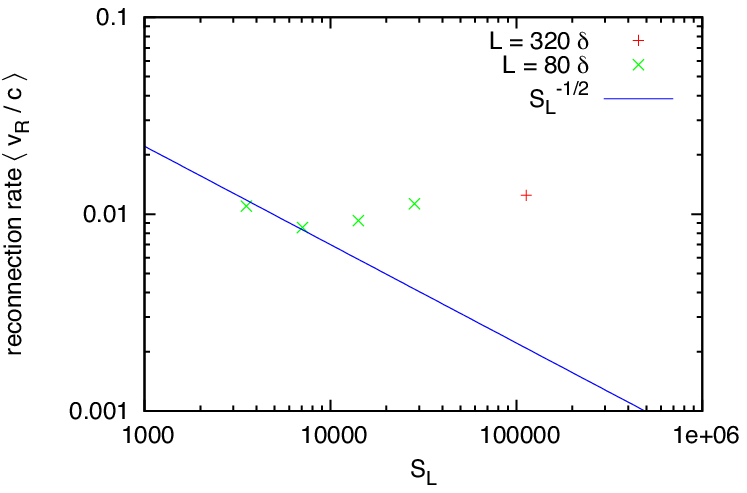}
  \caption{The plot of the time averaged reconnection rate $\langle v_R / c_A \rangle$ over the statistical equilibrium region with respect to the Lundquist number $S_L$. 
           Top: The strongly magnetized case: $\sigma_{in}= 14$. Bottom: The weakly magnetized case: $\sigma_{in}= 0.14$. 
          }
  \label{fig:5.2.4}
\end{figure}

Fig. \ref{fig:5.2.4} are the time-averaged reconnection rate $\langle v_R / c_A \rangle$ as a function of the Lundquist number $S_L$
\footnote{
Our numerical code includes the following numerical dissipation, $\eta_{num} \sim 0.03 c \Delta$, 
where $\Delta$ is the mesh size. 
This means 
our numerical code can calculate accurately problems with the Lundquist number up to $S_{num} = L c_A / \eta_{num} \sim 20 N c_A / c$ 
where $N$ is the mesh number along the current sheet. 
As explained in Sec. 3, 
we use the mesh number $N = 8192$ along the current sheet, 
our calculation has sufficient accuracy up to $S_L \sim 3 \times 10^5$. 
}
. 
The top panel is the relativistically strong magnetic field case, $\sigma_{in}= 14$, 
and the bottom panel is the non-relativistic magnetic field case, $\sigma_{in}= 0.14$. 
%We calculate the time average between the initial time and a saturation time 
%when the plasmoid-chain reaches a statistical equilibrium state and the averaged reconnection rate saturates. 
We calculate the time average of the reconnection rate curves over the plateau region 
where the plasmoid-chain reaches a statistical equilibrium state. % and the averaged reconnection rate saturates. 
As in the non-relativistic case, %(Loureiro et al. 2012 ...), 
we find that 
the reconnection rate becomes independent of the Lundquist number 
when it is larger than a critical value $S_c$. %$S_c \sim 10^4$. 
For small Lundquist numbers, % region, $S_L < S_c$, 
we find the Sweet-Parker sheet dependence $S_L^{-1/2}$ of the reconnection rate 
predicted by~\citet{2003ApJ...589..893L} and \citet{2005MNRAS.358..113L} (see also Eq. (\ref{eq:a11})). 
In our calculations, 
the critical value of the weakly magnetized case is $S_c \sim 10^4$, 
which is similar to the value indicated in the non-relativistic work; 
on the other hand, 
in the strongly magnetized case the critical value is $S_c \sim 2-3 \times 10^3$, 
which is a little less than that of the weak magnetic field case. 
This can be explained as follows. 
After generating plasmoids, 
the current sheet between the plasmoids will become a Sweet-Parker current sheet. 
In this case, 
the sheet thickness can be obtain by Eq. (\ref{eq:a4}). 
If we assume the reconnection jet velocity is the Alfv\'en velocity, 
%the sheet thickness can be written as, $\delta = \eta / 2 \sigma_{in} c_A$, 
the sheet thickness can be written as, $\delta = L / \sqrt{2 \sigma_{in} S_L}$, 
where we used Eq. (\ref{eq:a23}) to estimate $v_{in}$. 
This means the sheet thickness decreases with increasing the magnetic field strength. 
On the other hand, 
the growth time of the tearing instability is, $\sim \sqrt{\delta^3 / \eta c_A}$. 
Using these two expressions, 
the growth time of the tearing instability of the secondary current sheet is 
\begin{equation}
  \label{eq:5.3.2}
%  \tau_{tearing,2nd} \sim \frac{\eta}{c_A^2} \sqrt{\frac{1}{8 \sigma_{in}^3}}
  \tau_{tearing,2nd} \sim \frac{\tau_{A,L}}{(2 \sigma_{in})^{3/4} S_L^{1/4}} \propto \sigma_{in}^{- 3/4} c_A^{-5/4}
  .
\end{equation}
This means as the magnetic field strength becomes strong, the secondary tearing instability grows faster 
and the plasmoid instability occurs much easier, 
especially along the reconnection jet resulted from the initially triggered plasmoid. 
Similarly, 
using the characteristic wavelength of the tearing instability, $\lambda_{tearing} \sim \delta [\delta c_A / \eta]^{1/4}$, 
the characteristic wavelength of the secondary tearing instability can be obtained as: 
\begin{equation}
  \label{eq:5.3.3}
%  \lambda_{tearing,2nd} \sim \frac{\eta}{2 \sigma_{in} c_A} \frac{1}{[2 \sigma_{in}]^{1/4}}
  \lambda_{tearing,2nd} \sim L / [(2 \sigma_{in})^{5/8} S_L^{3/8}] \propto \sigma_{in}^{-5/8} c_A^{-3/8}
  .
\end{equation}
This also indicates that 
the plasmoid instability evolves more easily as the background magnetization parameter becomes larger. 
Note that 
Eq. (\ref{eq:5.3.3}) means that 
a background plasma with larger magnetization parameter demands a smaller Lundquist number with respect to the sheet length for the plasmoid instability 
due to the smaller characteristic wavelength of the secondary tearing instability. 
This also supports the results shown in Fig. \ref{fig:5.2.4}, 
which indicates that the critical Lundquist number becomes smaller as the magnetization parameter of the background plasma becomes larger. 
%In addition, 
%the induction equation indicates that 
%the current sheet thickness grows as $\delta \propto \sqrt{\eta t}$ due to the magnetic field diffusion by the resistivity $\eta$. 

As pointed out by \citet{2010PhRvL.105w5002U}, 
the reconnection rate of the plasmoid-chain can be written as, $v_R / c_A \sim 1 / \sqrt{S_c}$, using the relation of the Sweet-Parker sheet. 
If we use the above critical values, $S_c = 3 \times 10^3$, in the strongly magnetized case, 
the reconnection rate is $\sim 0.02 c_A$, 
which agrees with the values indicated in the top panel of Fig. \ref{fig:5.2.1}. 

\subsection{\label{sec:sec5.3}Evolution of Plasmoid Structure}

\begin{figure}[h]
 \centering
  \includegraphics[width=3.5cm,clip]{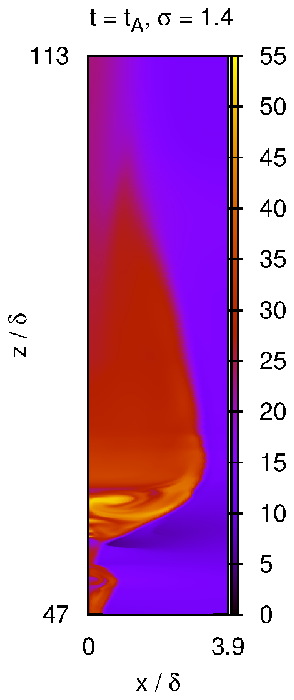}
  \includegraphics[width=3.5cm,clip]{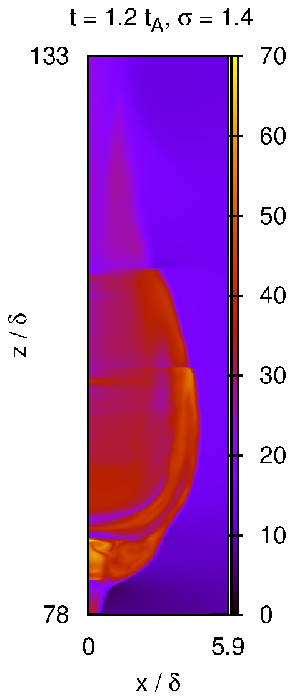}
  \caption{Snapshots of the density profile of the initially triggered plasmoid in the case of $\sigma_{in}= 1.4$. 
           The left panel is at $t = t_A$ and the right panel is at $t = 1.2 t_A$. 
          }
  \label{fig:5.3.1}
\end{figure}

\begin{figure}[t]
 \centering
  \includegraphics[width=3.5cm,clip]{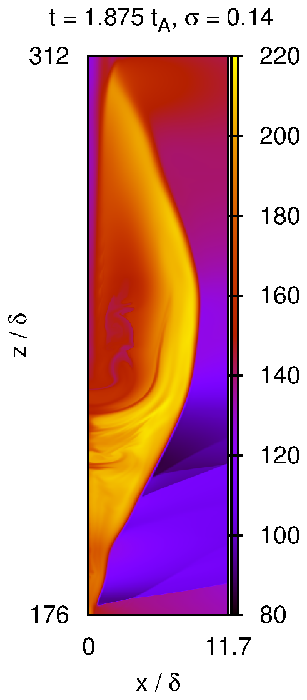}
  \includegraphics[width=3.5cm,clip]{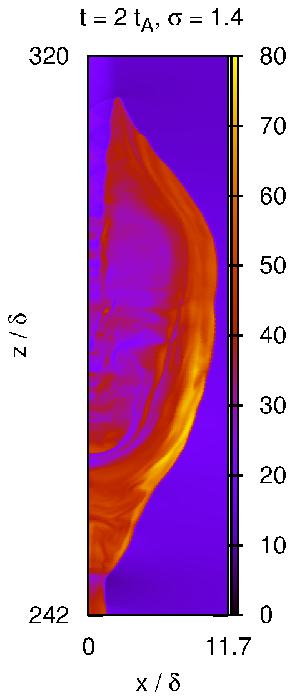}
  \caption{Snapshots of the density profile of the initially triggered plasmoid. 
           The left panel is at $t = 1.875 t_A$ with weakly magnetized case, $\sigma_{in}= 0.14$, 
           and the right panel is at $t = 2 t_A$ with strongly magnetized case, $\sigma_{in}= 1.4$. 
          }
  \label{fig:5.3.2}
\end{figure}

Fig. \ref{fig:5.1.1} shows that 
the aspect ratio of plasmoids takes different values, 
depending on the magnetization parameter $\sigma_{in}$; 
the aspect ratio seems to take a smaller value  
as the magnetization parameter $\sigma_{in}$ increases. 
This can be explained as follows. 
Left panel of Fig. \ref{fig:5.3.1} is the density profile of a plasmoid at $t = t_A$. 
This figure shows its aspect ratio is about $14:1$ 
and the inner structure of the plasmoid is very similar to that of the Petschek reconnection case 
which was investigated by~\citet{2011PhPl...18b2105Z}. 
Right panel of Fig. \ref{fig:5.3.1} is the density profile of the same plasmoid at $t = 1.2 t_A$. 
We find that 
%the plasmoid size in z-direction shrinks by slow shocks. 
the plasmoid size in z-direction shrinks because of the appearance of slow shocks. 
%These slow shocks are generated by two processes basically. 
%The first process is the steepning of slow waves 
%which are generated by collisions to another plasmoids. 
%In the case of Figs. \ref{fig:5.3.1}, 
%these slow waves are generated the plasmoid at $y \sim 48 \delta$ in the left panel. 
%The second process is the evolution of the corrugation instability of the slow shock~(J.M.Stone \& M. Edelman 1995). 
%The non-relativistic investigation shows that 
%strong oblique slow shocks are unconditionally unstable to perturbations perpendicular to shock fronts 
%and the growth rate of the corrugation is proportional to the square of the Alfv\'enic Mach number $M_{A0} \equiv v_0 / c_A$ 
%where $v_0$ is the slow shock velocity. 
These shocks are generated by the steepening of slow waves 
which are induced by collisions with other plasmoids. 
In the example shown in Fig. \ref{fig:5.3.1}, 
slow waves are generated by the collision to the plasmoid at $z \sim 48 \delta$ in the left panel. 
As these slow shocks propagate across the plasmoid, 
the upstream plasma in the plasmoid is compressed 
and the plasmoid size shrinks in z-direction. 
Fig. \ref{fig:5.3.2} shows the density configuration of the plasmoid triggered by the initial perturbation 
of runs B1, B2 at a time just before it escapes from the numerical domain. 
In run B1, $\sigma_{in}= 0.14$, 
we find the aspect ratio of the plasmoid keeps its initial value, approximately $14:1$. 
This is because 
in run B1 the plasmoid instability does not grow sufficiently as explained in the previous sections 
and the largest plasmoid does not experience a collision with a smaller plasmoid. 
On the other hand, 
in runs B2 
many collision with smaller plasmoids reduce the aspect ratio of the plasmoids to about $6:1$. 
In our calculations, 
the aspect ratio does not show any rapid time evolution 
after it reaches the above ratio, $6:1$. 
Although we cannot be certain that this ratio is the final state, 
it seems that the aspect ratio depends very weakly on time. 
Finally, 
we cannot find any strong dependence of the above aspect ratio on the magnetization parameter $\sigma$. 

\section{\label{sec:sec6.1}Trajectory of X and O-Points}
\begin{figure}[h]
 \centering
  \includegraphics[width=10cm,clip]{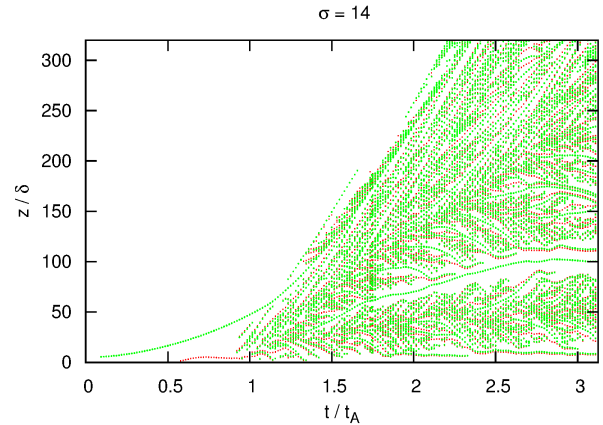}
  \caption{Trajectories of X and O points of the run B3 along the current sheet. 
           Green points are the O-points and red points are the X-points. 
          }
  \label{fig:6.1.1}
\end{figure}

To understand the physical nature of the plasmoid-chain, 
it is helpful to trace trajectories of the X and O points 
that are the magnetic null points: $B_x = B_z = 0$. %, 
%and only the guide field is finite. 
%An X-point is a point where magnetic reconnection occurs 
%and an O-point is usually considered to be the location of a plasmoid. 
At X-points, 
the magnetic configuration around them is the X-type 
and those are points where magnetic reconnection occurs; 
at O-points, 
the magnetic configuration around them is the O-type 
and they are usually equivalent to the location of a plasmoid. 
Fig. \ref{fig:6.1.1} is a plot of the trajectories of X and O points of run B3. 
This figure shows that 
in the initial phase there is only one O-point 
which is generated by the initial perturbation at the origin. 
Around $t = t_A$, 
small plasmoids start to develop behind the initial O-point 
and the number of O-points and X-points gradually increases with time. 
Around $t = 2.2 t_A$, 
the initial plasmoid reaches the boundary of the numerical domain and escapes from the domain. 
After that, the current sheet is filled with X and O points, 
the plasmoid-chain is fully evolved. 
This is consistent with the temporal evolution of the reconnection rate shown in Fig. \ref{fig:5.2.1}. 
Fig. \ref{fig:6.1.1} shows that 
most points, particularly those close to the initial plasmoid, move steadily towards larger $z$. 
Their velocity is approximately $0.8 c$. 
%Note that most X and O points move upwards until $800 t_A$. 
%Afterwards, they, however, gradually start to move in both directions along z-direction  
%as the plasmoid-chain evolves, 
%and it reaches a statistical equilibrium state at $t = 1000 t_A$. 
Note that X and O points 
which are not close to the initial plasmoid 
gradually start to move in both directions along z-direction. 
This region is confined around the origin initially  
and expands with time as the plasmoid-chain evolves. 
Finally this region covers all the simulation domain 
and the plasmoid-chain reaches a statistical equilibrium state around $t = 3 t_A$. 

Concerning X-points, 
we find that 
they are located near the midpoint between two O-points as is expected 
since they are generated by the tearing instability which ejects two plasmoids away from the X-point. 
Fig. \ref{fig:6.1.1} shows that 
many X-points move along the current sheet and most of them disappear after a short time 
due to the merger of two neighboring plasmoids or the collapse of X-points~\citep{2005PhRvL..95w5003L}. 
In addition, 
we find that 
X-points that move in a way similar to that of the nearest plasmoid 
as reported by \citet{2011ApJ...737...24B}. 
%which is induced by the motion of the nearest plasmoids. 
Since X-points are considered to play an important role for the particle acceleration, 
their dynamical time along the current sheet will impose an upper limit on the acceleration time. 
For example, 
if we consider a current sheet with a plasmoid-chain in statistical equilibrium, 
with a critical Lundquist number $S_c$, 
the sheet length between them can be estimated as: $L_c \sim S_c \eta / c_A$; 
the dynamical time can be estimated as 
\begin{equation}
  \label{eq:6.1.1}
  t_{acc} \sim L_c / c_A \sim S_c \eta / c_A^2
  ,
\end{equation}
and direct acceleration by the electric field at X-points will be limited by this time scale. 
Using our parameters, 
the value of the acceleration time is $t_{acc} \sim 1.5 \times 10^{-2} t_A$. 
Note that Fig. \ref{fig:6.1.1} includes X-points whose lifetime is much longer than the above value. 
Their typical lifetime is about $1.5 \times 10^{-1} t_A$ 
and some of them survive for a much longer time. 
Fig. \ref{fig:6.1.1} indicates that 
they accompany large plasmoids 
which have somewhat large spaces around them. 

Note that 
sometimes large spaces appear in the current sheet in Fig. \ref{fig:6.1.1}, 
such as $z = 0$ or $z = 100 \delta$. 
This is due to the ``\textit{monster plasmoids}'' 
which result from the merger of many smaller plasmoids. 
In particular, 
the monster plasmoid at $z = 100 \delta$ around $t = 3 t_A$ shows interesting behavior. 
In the initial phase, 
it behaves in nearly the same as other plasmoids. 
In the later phase, 
its inertia becomes much larger than that of surrounding plasmoids, 
and its dynamics starts to resemble Brownian motion, 
since it moves stochastically around an average trajectory 
that has a low velocity. 
%However, as it evolves, 
%it starts to move at a slower velocity. 
%This is because as it experience many mergers, 
%its inertia, $\rho c^2 + p / (\Gamma - 1)$, also increases. 
%In the later phase, 
%its inertia becomes much larger than the surrounding plasmoids, 
%and the collision process becomes the Fokker-Planck like process, 
%so that it moves stochastically around its average coordinate at a small velocity. 

\begin{figure}[h]
 \centering
  \includegraphics[width=7cm,clip]{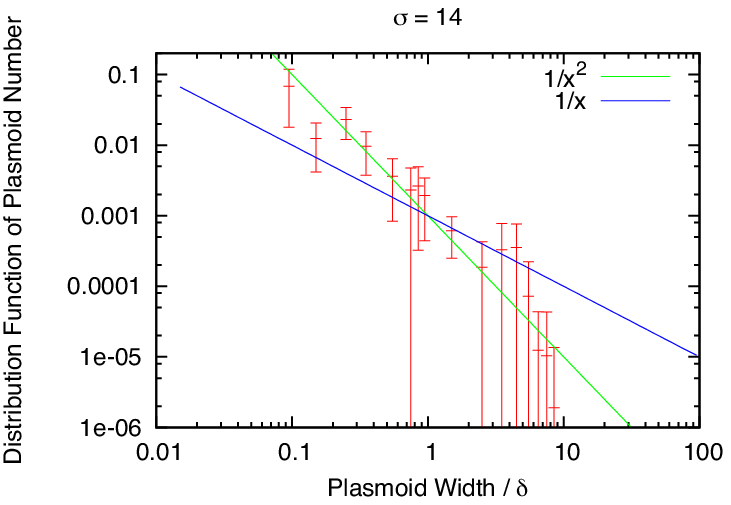}
  \includegraphics[width=7cm,clip]{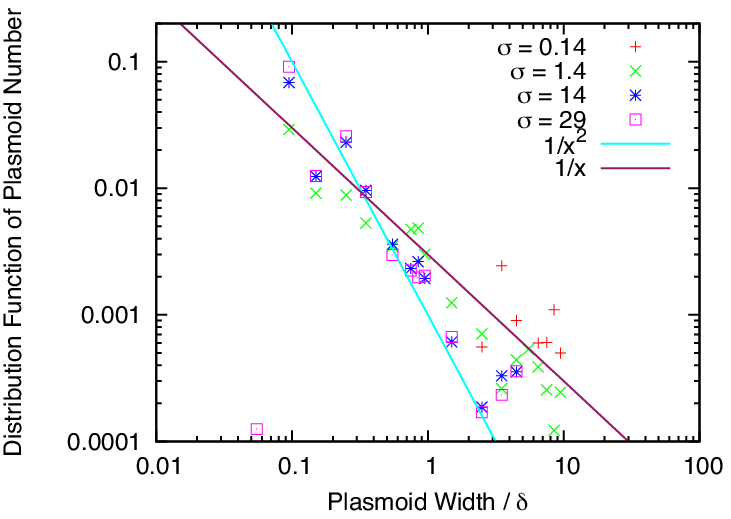}
  \caption{The time-averaged distribution function of plasmoid size perpendicular to the current sheet. 
           The distribution functions are averaged over between $t = t_A$ and $t = 2.2 t_A$. 
%           of run B1-B4. 
           Top: The distribution of run B3 with error bars. 
           Bottom: The distributions of run B1-B4. 
          }
  \label{fig:6.2.1}
\end{figure}

\section{\label{sec:sec6.2}Plasmoid Size Distribution}

As mentioned in the previous sections, 
the evolution of a plasmoid induces a secondary tearing instability, 
and generates small plasmoids behind it; 
the small plasmoids in turn induce more tearing instabilities, 
and as a result the current sheet evolves into the plasmoid-chain. 
%Since the distribution of plasmoid size can give us a lot of fruitful implications for high-energy astrophysical phenomena, 
Since the distribution of plasmoid size is potentially important for high-energy astrophysical phenomena, 
%for the stricture of Y-point current sheet and the relation to the gamma ray pulse of the Crab pulsar (Lyubarsky 2012), 
we investigate this using our numerical results. 
%Because of the self-similar nature of the MHD equation, 
%the plasmoid-chain shows a power law distribution of the plasmoid width. 

The statistical behavior of the plasmoid-chain was investigated in \citep{2010PhRvL.105w5002U,2010PhPl...17a0702F,2011JGRA..116.9226F,2012PhPl...19d2303L,2012PhRvL.109z5002H,2013arXiv1301.0331H}.  
In those papers, 
the authors discuss the time evolution of the distribution function of the plasmoid-chain using the following model kinetic equation: 
\begin{equation}
  \label{eq:6.2.1}
  \frac{\partial f}{\partial t} + \alpha \frac{\partial f}{\partial \Psi} = \zeta \delta(\Psi) - \frac{f N}{\tau_A} - \frac{f}{\tau_A}
  ,
\end{equation}
where $f(\Psi)$ is the distribution function, 
$\Psi$ is the magnetic flux of a plasmoid, 
$N(\Psi) \equiv \int^{\infty}_{\Psi} f(\Psi') d \Psi'$ is the cumulative distribution function, 
$\alpha \sim B_0 c_A / \sqrt{S_c}$ is the plasmoid growing rate of a plasmoid, 
$\tau_A \sim L / c_A$ is the Alfv\'en crossing time of a plasmoid across the plasmoid-chain with scale $L$, 
and $\zeta$ is the magnitude of the source of plasmoids. 
Thus, 
the second term on the left-hand side describes the growth of plasmoids; 
the first term on the right-hand side is the source of plasmoids; 
the second term is the loss of plasmoids due to mergers with larger plasmoids; 
the third term is the advection loss. 
Some analytical steady state solutions of Eq. (\ref{eq:6.2.1}) in large $\Psi$ region can be obtained as follows. 
When the loss of plasmoids is mainly by advection, $N \ll 1$, 
we obtain $f \propto \exp [ - \Psi / \alpha \tau_A]$; 
when the loss of plasmoids is mainly by plasmoid merger, $N \gg 1$, 
we obtain $f \sim 2 \alpha \tau_A \Psi^{-2}$. 
%The authors obtained the distribution of the flux function and the width of plasmoids in the plasmoid-chain. 
%In both cases, 
%they predicted power law distributions with index $-2$. 
In this derivation, 
we assumed the speed of plasmoids is of the order of $c_A$, corresponding to an assumption of the plasmoid crossing time as $\tau_A \sim L/c_A$. 
Recently, 
\citet{2013arXiv1301.0331H} showed that 
dropping this assumption allows a solution $f \propto \Psi^{-1}$. 
%the power law index of the distribution function can be $-1$ 
%if we do not assume the above assumption of the plasmoid velocity. 
Since the magnetic flux can be expressed as: $\Psi \sim B_0 w$ 
where $w$ is the plasmoid size perpendicular to their current sheet~\citep{2010PhRvL.105w5002U}, 
the above distribution function of the magnetic flux can be used to find the plasmoid size distribution. 

\begin{figure}[t]
 \centering
  \includegraphics[width=7cm,clip]{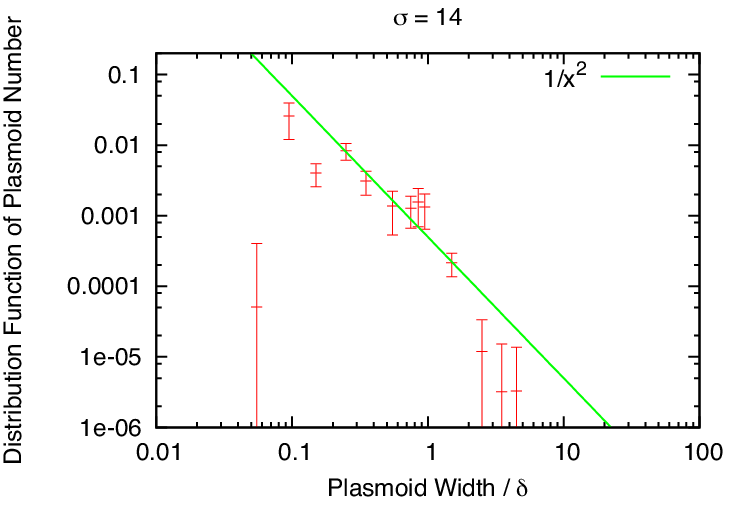}
  \includegraphics[width=7cm,clip]{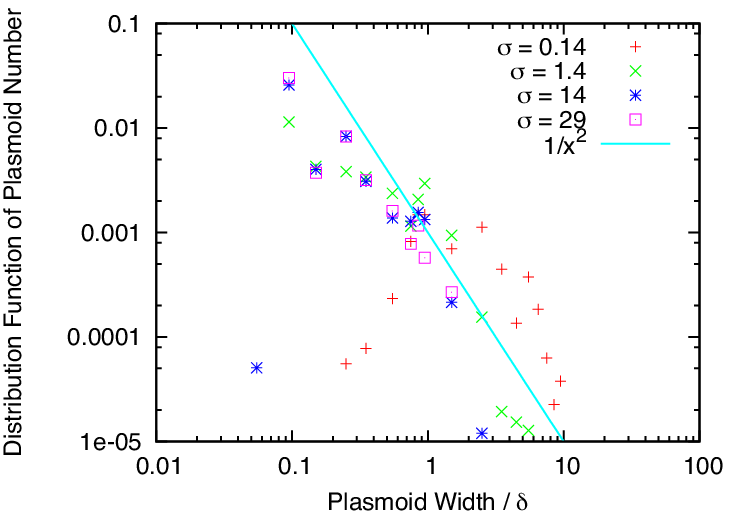}
  \caption{The time-averaged distribution of plasmoid size perpendicular to the current sheet. 
           The distribution functions are averaged over after $t = 2.2 t_A$. 
%           of run B1-B4. 
           Top: The distribution of run B3 with error bars. 
           Bottom: The distributions of run B1-B4. 
          }
  \label{fig:6.2.2}
\end{figure}

The top panel of Fig. \ref{fig:6.2.1} is the time averaged distribution of the plasmoid size of run B3 with error bar. 
The time-average is taken over between $t = t_A$ and $t = 2.2 t_A$ 
each of which is equivalent to the starting time of plasmoid instability 
and the escaping time of the initially triggered plasmoid, respectively, 
as indicated in Fig. \ref{fig:5.2.1} of Sec. \ref{sec:sec5.2}. 
%The top panel of Fig. \ref{fig:6.2.1} is a snapshot of the distribution of the plasmoid size of run B3 
%when the initially generated plasmoid reached the upper limit of the numerical domain. 
This figure shows that 
the plasmoid size distribution is consistent with a size distribution of power law index $-2$ %in the large scale region, 
for small plasmoids in the range $[0.1 \delta, \delta]$ of the plasmoid width, 
as predicted by previous works for the non-relativistic case. 
From the above discussion, 
this means that 
the plasmoid loss is mainly due to plasmoid mergers. 
This is a natural consequence because 
we consider the distribution at the escape time of the initially triggered plasmoid 
and any plasmoids cannot escape from the plasmoid-chain at that time due to the presence of the initially triggered plasmoid. 
%In the larger region, 
In addition, 
larger plasmoids, around $\delta < w < 5 \delta$, deviate from the power law index $-2$ 
%the distribution deviates from the power index $-2$ 
%and it becomes approximately $-1$. 
and tend to an index of $-1$. 
This indicates that 
%the velocity of small plasmoids is of order of the Alfv\'en velocity 
%and that of large plasmoids deviates from the Alfv\'en velocity. 
the velocity of large plasmoids deviates from the Alfv\'en velocity. 
This is because 
the large plasmoids have large inertia, 
which reduces their velocity, 
as in the case of the monster plasmoid. 
%This is due to the 
%the number of plasmoids increases with size. 
%This may be because 
%the resolution of our calculations is not sufficient for reproducing the plasmoid evolution in small scale. 
Note that the distribution function of the largest plasmoid size region, around $w > 5 \delta$, drops rapidly 
and clearly deviates from power law. 
This is because the number of plasmoids is too small to show statistically sufficient results. 
This can also be seen from the large error bar of this region. 

In the bottom panel of Fig. \ref{fig:6.2.1}, 
we plot the plasmoid size distribution of runs B1-B4. 
We find that 
the distribution of the strong magnetic field case, run B4, shows very similar behavior to the run B3; 
it becomes a power law with index $-2$ in the range $[0.1 \delta, \delta]$ of the plasmoid width %in the large scale region. 
and $-1$ for larger plasmoids. % width region. 
In the weak magnetic field cases, runs B1,B2, 
the distribution also has an index of $-1$ in the range of larger plasmoid. % width region; 
However, the distribution of the smaller plasmoid size region seems to have an index of $-1$, too. 
We consider this is because small plasmoids are not sufficiently evolved in these runs 
to show a clear size dependence. 

In Fig. \ref{fig:6.2.2}, 
we plot the time-averaged distribution functions 
after the initially triggered plasmoids escaped: $t > 2.2 t_A$. 
The top panel of Fig. \ref{fig:6.2.2} is the time averaged distribution of the plasmoid size of run B3 with error bar. 
In the small plasmoid region, 
the distribution function has an index of $-2$, 
similarly to the previous case. 
However, 
the distribution function of the larger plasmoid region, $w > \delta$, drops rapidly 
and clearly cannot be approximated by the power law. 
We consider 
this is due to the effect of the plasmoid loss by advection. 
Since the initially triggered plasmoid already escaped from the simulation domain in this case, 
the plasmoids can freely escape from the domain 
and this results in the exponential decay of the distribution function, 
as indicated by the above discussion using the kinetic equation. 

The bottom panel of Fig. \ref{fig:6.2.2} is the plot of the plasmoid size distribution of runs B1-B4. 
The behavior of the distribution functions in small plasmoid region, $w < \delta$, is very similar to Fig. \ref{fig:6.2.1} 
but that in large plasmoid region also show rapid decay, 
similarly to the strongly magnetized case, $\sigma = 14$. 
Note that the distribution function of the weakly magnetized case, $\sigma = 0.14$, seems to be a power law 
in large plasmoid region, $w > \delta$. 
Unfortunately, 
our data does not have sufficiently large number of plasmoids in this region, 
so that we cannot conclude that this is a statistically correct result. 

%they become the power law with index $-1$ in the large plasmoid width region; 
%in the small plasmoid width region, 
%they do not show clear dependence on the plasmoid width. 
%We consider this is due to the insufficient evolution of small plasmoids. 

%large plasmoid interveen advection escape => power law
%
%low beta => small plasmoids => v-v' ~ vA => -2
%            large plasmoids => v-v' ~/ vA => -1
%
%high beta => the number of plasmoids is small and a lot of large plasmoids => -1

\section{\label{sec:sec7}Summary}

In this paper, 
we investigated the evolution of the plasmoid-chain in a high-$\sigma$ plasma. 
We modeled the relativistic current sheet with cold background plasma using the relativistic resistive magnetohydrodynamic approximation, 
and solved its temporal evolution numerically. 
We performed various calculations using different magnetization parameters of the background plasma from $\sigma_{in}= 0.14$ to $\sigma_{in}= 29$ 
and different Lundquist numbers with respect to the sheet length from $S_L \sim 10^3$ to $S_L \sim 10^5$. 
The numerical results show that 
the initially induced plasmoid triggers a secondary tearing instability 
and the current sheet is gradually filled with many plasmoids, that is, it evolves into a plasmoid-chain, 
as predicted by non-relativistic work. 
As expected, 
this plasmoid instability enhances the reconnection rate, 
which grows until the initially triggered plasmoid escapes from the simulation domain, 
reaching up to $\sim 0.05 c_A$. 
Subsequently, the plasmoid-chain reaches a statistically equilibrium state, 
and the temporally averaged reconnection rate in a steady state becomes $\sim 0.03 c_A$. 
Since the maximum value of the Alfv\'en velocity is the light velocity $c$, 
our numerical calculation indicates the maximum reconnection rate of the plasmoid-chain is $0.03 c$. 
In our calculations, 
the evolution of the reconnection rate shows similar behavior 
in strongly magnetized cases: $\sigma_{in} > 1$. 
Although the weakly magnetized case, $\sigma_{in}= 0.14$, shows different behavior, 
we consider this is due to the larger wavelength of the secondary plasmoid instability indicated by Eq. (\ref{eq:5.3.3}). 
Note that the above critical value is much smaller than that obtained recently by \citet{2011MNRAS.418.1004Z}, 
who found $S_c \sim 10^8$. 
We believe this difference comes from their assumption of a relativistically hot background plasma. 
A high temperature reduces the magnetization parameter $\sigma_{in} = B_0^2 / 4 \pi \rho_0 h_0 \gamma_0^2$ 
and the critical Lundquist number becomes large when $\sigma$-parameter is small, as shown in Sec. \ref{sec:sec5.2}. 

We also investigated the behavior of O-points and X-points. 
In our simulations, 
the initial perturbation is confined to the origin. 
The triggered plasmoid shrinks the current sheet behind of it, 
inducing secondary tearing instabilities. 
%In this paper, 
%we gave initial perturbation only at the origin. 
%The initially triggered plasmoid shrinks the current sheet behind of it 
%and induces secondary tearing instabilities. 
Those O and X points that are close to the triggered plasmoid move in the same direction, 
but the other points start to move in both directions along the sheet, 
reflecting the final state of statistical equilibrium of the plasmoid-chain. 
Most X and O points disappear by merging, 
which limits their lifetime, 
and therefore, limits the time 
for which particles can be accelerated by the electric field at such points. 
%Though O and X points close to the initially triggered plasmoids move in the same direction as the initial plasmoid, 
%other points start to move in both directions 
%and finally it reaches a statistically equilibrium state of the plasmoid-chain. 
%During the evolution of the plasmoid-chain, 
%we observed many mergers of X and O points. 
%Most X and O points disappeared due to the mergers 
%and it limits their lifetime. 
%acceleration
%We estimate this lifetime using the relation of the plasmoid-chain 
%and it can be used for the typical upper limit of the acceleration time of particles by the electric field. 
We estimate this lifetime using the parameters of the plasmoid-chain. 
%and it can be used for the typical upper limit of the acceleration time of particles by the electric field. 
As predicted for the non-relativistic case, 
we noted the appearance of the ``\textit{monster plasmoids}''. 
Interestingly, 
our calculations show that 
monster plasmoids slow down as they evolve, 
because of their increasing inertia. 
Ultimately, they display Brownian like motion around fixed points. 
%monster plasmoid and its motion

Finally, 
we investigated the plasmoid size distribution. 
Our numerical results show that 
in strongly magnetized cases 
the distribution becomes power law with index $-2$ in the small plasmoid region 
and $-1$ in the large plasmoid region 
before the initially triggered plasmoid escapes. 
This indicates that 
the plasmoid loss is mainly due to mergers; 
the plasmoid velocity is of order of the Alfv\'en velocity in the small plasmoid region, 
but is lower in the large plasmoid region. 
This is because the plasmoid inertia increases with increasing size, 
preventing large plasmoids from moving at the Alfv\'en speed. 
%and it prevents large plasmoids moving at the Alfv\'en velocity. 
%and takes different value from the Alfv\'en velocity in the large plasmoid region. 
%We consider 
%this is because the inertia of plasmoids increases as the plasmoid size becomes larger, 
%and it prevents large plasmoids moving at the Alfv\'en velocity. 
%plasmoid size distribution
%power law  -> large -1, small -2     agree with theory
After the escape of the initial plasmoid, 
the distribution function in large plasmoid region shows exponential decay 
because of the free advective escape of plasmoids from the domain. 

% very important Y-point     
% resistivity   = depends its surcumstans  =>  in future work, we will investigates each phenomena appropriately, like  photon dragging and so on
%As shown in (Barkov \& Komissarov 2007), 
Magnetic reconnection is one of the most efficient mechanisms of magnetic field dissipation, 
and is expected to play an important role in many astrophysical phenomena. 
As shown in this paper, 
once the tearing instability evolves and generates plasmoids, 
the plasmoid-chain always evolves in the current sheet between them 
if the Lundquist number of the current sheets is beyond the critical value, 
especially in the Poynting-dominated plasma. 
Since plasmoids are associated with high temperature plasma and accelerated particles, 
they can be used to explain intermittent observational signals from high energy astrophysical objects, 
such as pulsed emission from the Y-point of the Crab pulsar magnetosphere \citep{2012arXiv1210.3346U} 
and multi-timescale TeV flares in blazars~\citep{2013MNRAS.431..355G}. 
In this paper, 
we assumed a constant resistivity and used an approximate equation of state corresponding to a relativistic, adiabatic gas. 
Nevertheless, we believe our results revealed general properties of plasmoid-chain in a Poynting dominant background plasma. 
%we should use appropriate form of them to apply for each phenomena. 
%The plasma state, however, are still unclear in many high astrophysical phenomena, 
%so they are our future works. 

\acknowledgments
%We would like to thank referees for many fruitful comments. 
We would like to thank John Kirk, Iwona Mochol, Simone Giacche, Seiji Zenitani, Keizo Fujimoto, Takaaki Yokoyama and Tsuyoshi Inoue 
for many fruitful comments and discussions. 
We also would like to thank our referees for a lot of fruitful comments on our paper. 
%We also would like to thank Jennifer M. Stone for improving the English. 
Numerical computations were carried out 
on SR16000 at YITP in Kyoto University. 
Calculations were also carried out on the Cray XT4 
at Center for Computational Astrophysics, CfCA, of National Astronomical Observatory of Japan.
This work is supported by Max-Planck-Institut f\"ur Kernphysik 
and the Postdoctoral Fellowships for Research Abroad program by the Japan Society for the Promotion of Science No. 20130253 (M. T.). 
%This work is supported by Grant-in-aids from the Ministry of Education, Culture, Sports, Science, and Technology (MEXT) of Japan, 
%No.22$\cdot$3369 and No. 23740154 (T. I.). 

%SR16000'à

\appendix
\section{\label{sec:secA2}Relativistic Sweet-Parker Current Sheet}
\begin{figure}[h]
 \centering
  \includegraphics[width=6cm,clip]{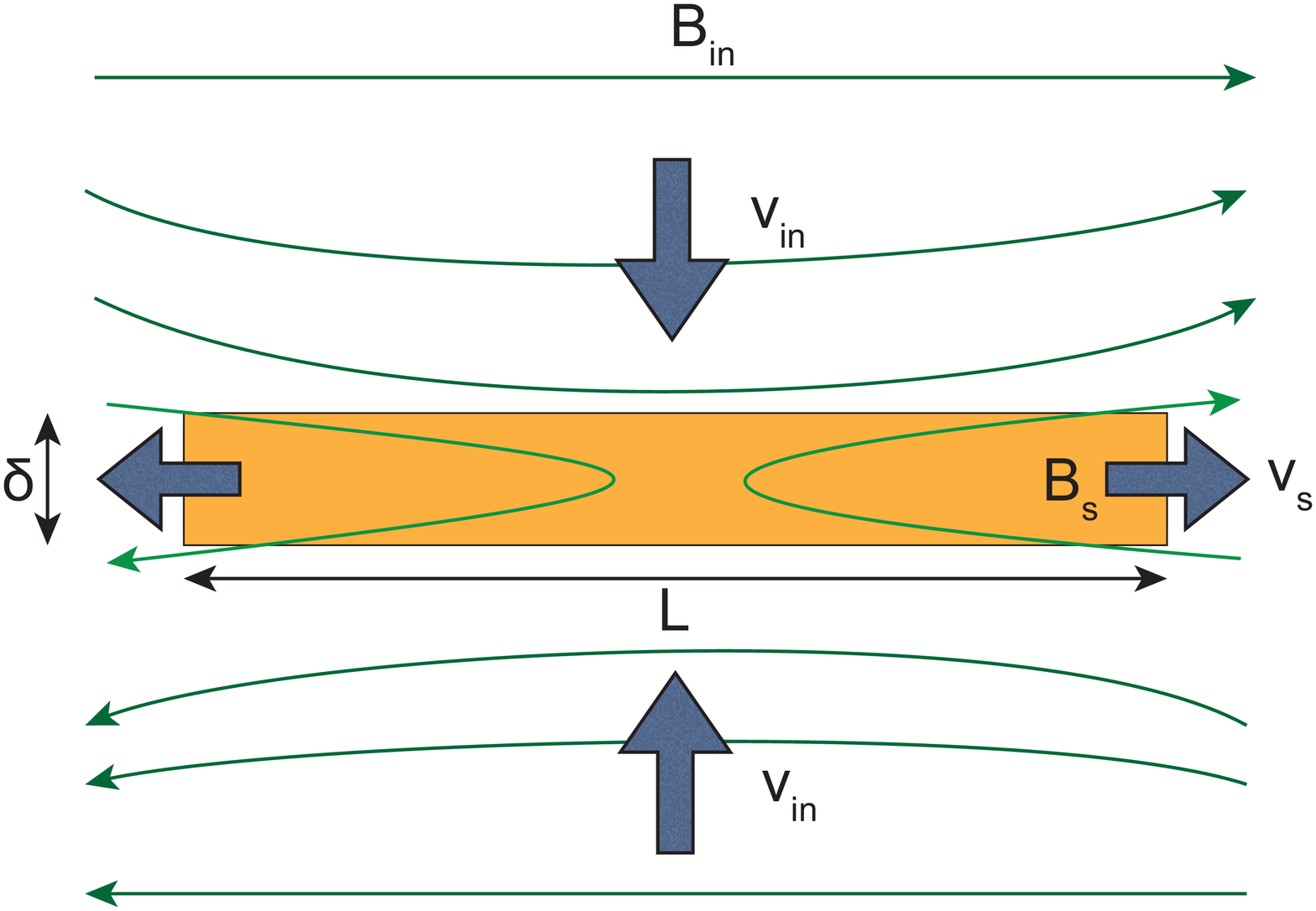}
  \caption{A schematic picture of the Sweet-Parker current sheet. 
          }
  \label{fig:1}
\end{figure}

In this appendix, 
we derive the reconnection rate of the relativistic Sweet-Parker current sheet. 
The basic relations have already been presented by several authors~\citep{1994PhRvL..72..494B,2003ApJ...589..893L,2005MNRAS.358..113L}. 
%In this Appendix, 
Here, 
we clarify the dependence on the external pressure, 
following the non-relativistic approach of~\citep{2000mare.book.....P}. 
%Basic relations of the relativistic Sweet-Parker sheet were already obtained, for example, in \citep{1994PhRvL..72..494B,2003ApJ...589..893L,2005MNRAS.358..113L}. 
%%In this Appendix, 
%We will present the relation of relativistic Sweet-Parker sheet 
%with clarifying its dependence on the outside pressure 
%following the non-relativistic work \citep{2000mare.book.....P}. 
A schematic picture of the Sweet-Parker current sheet is shown in Fig. \ref{fig:1}. 

We assume a steady state plasma 
which can be described well by the relativistic magnetohydrodynamic approximation other than around the X-point. 
We also assume that 
the plasma is homogeneous in the y-direction. 
The background magnetic field of the inflow region is ${\bf B}_{in} = B_{in} {\bf e}_x$ if $z > \delta / 2$ and ${\bf B}_{in} = - B_{in} {\bf e}_x$ if $z < - \delta / 2$, 
and that in the sheet region is ${\bf B} = \epsilon {\bf B}_{in} \pm B_s {\bf e}_z$ 
where $|B_{in}| \gg |B_s|$ and $\epsilon$ is a very small constant. 
We assume that ${\bf B} = {\bf 0}$ at the X-point. 
In this case, 
the electric field $E_y$ is constant, 
and we can obtain the following relation: 
\begin{equation}
  \label{eq:a1}
  B_{in} v_{in} = B_s v_s
  ,
\end{equation}
where $v$ is the fluid 3-velocity. 
In addition, 
we can obtain the following relation at the X-point 
where the magnetic field and flow velocity is $0$: 
\begin{equation}
  \label{eq:a2}
  B_{in} v_{in} = \eta j
  ,
\end{equation}
where $\eta$ is the resistivity and $j$ is the current vector described by the Amp\'ere's law 
\begin{equation}
  \label{eq:a3}
  j \sim B_{in} / \delta
  .
\end{equation}
From Eqs. (\ref{eq:a2}) and (\ref{eq:a3}), 
the sheet thickness $\delta$ can be expressed as: 
\begin{equation}
  \label{eq:a4}
  \delta \sim \eta / v_{in}
  .
\end{equation}
The mass and the energy conservation equation are given by 
\begin{eqnarray}
  \label{eq:a5}
  \rho_{in} \gamma_{in} v_{in} L &=& \rho_s \gamma_s v_s \delta
  ,
  \\
  \label{eq:a6}
  (\rho_{in} h_{in} \gamma_{in}^2 + B_{in}^2) v_{in} L &=& (\rho_s h_s \gamma_s^2 + B_s^2) v_s \delta
  , 
\end{eqnarray}
where $\rho$ is the rest mass density, $\gamma$ is the Lorentz factor, 
$L$ is the curvature scale of the background magnetic field 
and $\delta$ is the current sheet thickness. 
$h = 1 + \Gamma / (\Gamma - 1) p / \rho$ is the specific entharpy of the ideal gas 
where $\Gamma = 4/3$ is the relativistic heat ratio and $p$ is the gas pressure. 
%$h$ is the specific entharpy. 
Here, 
we also assume the cold upstream plasma, $p_{in} = 0$; 
in the sheet region we assume a hot plasma $\rho_s \ll p_s$ 
whose pressure can be determined through the pressure equilibrium, $p_{s} = B_{in}^2 / 2 \gamma_{in}^2$. 
Then, the energy equation can be rewritten as 
\begin{equation}
  \label{eq:a7}
  \rho_{in} \gamma_{in}^2 (1 + \sigma_{in}) v_{in} L = \left[ \frac{2 B_{in}^2 \gamma_s^2}{\gamma_{in}^2} + B_s^2 \right] v_s \delta
  ,
\end{equation}
where $\sigma \equiv B^2 / \rho h \gamma^2$ is the magnetization parameter. 
Using Eqs. (\ref{eq:a1}), 
the above equation reduces to 
\begin{equation}
  \label{eq:a8}
  (1 + \sigma_{in}) \gamma_{in}^2 v_{in} = \left[ 2 \sigma_{in} \gamma_s^2 + \frac{\sigma_{in}}{v_s^2} \gamma_{in}^2 v_{in}^2 \right] v_s \frac{\delta}{L}
  .
\end{equation}
Using Eqs. (\ref{eq:a4}), 
\begin{equation}
  \label{eq:a9}
  (1 + \sigma_{in}) \gamma_{in}^2 v_{in}^2 \sim \left[ 2 \gamma_s^2 v_s^2 + \gamma_{in}^2 v_{in}^2 \right] \frac{\sigma_{in}}{S_l v_s}
  ,
\end{equation}
where $S_l \equiv L c / \eta$ is the Lundquist number using the light velocity as the characteristic velocity. 
From this equation, 
we can obtain the following relation between $v_{in}$ and $v_s$:
\begin{equation}
  \label{eq:a10}
  \gamma_{in} v_{in} \sim \sqrt{\frac{2}{S_l v_s - c_A^2}} \gamma_s v_s c_A
  ,
\end{equation}
where $c_A \equiv \sqrt{\sigma / (1 + \sigma)}$ is the Alfv\'en velocity. 
If we consider a plasma with high-Lundquist number $S_l \gg 1$, 
the above equation reduces to 
\begin{equation}
  \label{eq:a11}
  \gamma_{in} v_{in} \sim \sqrt{\frac{2}{S_l}} \gamma_s \sqrt{v_s} c_A
  .
\end{equation}
This equation shows that 
the inflow velocity, the reconnection rate, is inversely proportional to $\sqrt{S_l}$, 
which is the same conclusion as the non-relativistic Sweet-Parker current sheet model. 
To obtain an explicit solution of the upstream velocity, 
we have to add another equation to the above equations. 
Here, 
we consider the equation of motion along the x-direction. 
The relativistic hydrodynamical equation of motion in the current sheet is given by 
\begin{equation}
  \label{eq:a12}
  \frac{\rho_s h_s \gamma_s^2 v_s^2}{L} \sim j B_s - \frac{p_o - p_N}{L} \sim \frac{B_{in}}{\delta} B_s - \frac{p_o - p_N}{L}
  ,
\end{equation}
where 
we used Eq. (\ref{eq:a3}), 
$p_o$ is pressure at the edge of the current sheet, 
and $p_N \sim p_s$ is pressure at the X-point, respectively. 
Note that the scale $L$ is a characteristic scale length in the above equation, 
and it should be the curvature scale of the background magnetic field. 
Using the pressure equilibrium, 
the above equation reduces to 
\begin{equation}
  \label{eq:a12}
  2 \frac{B_{in}^2}{\gamma_{in}^2} \frac{\gamma_s^2 v_s^2}{L} \sim \frac{B_{in}}{\delta} B_s - \frac{p_o - p_N}{L}
  ,
\end{equation}
From the mass conservation equation Eq. (\ref{eq:a5}) and Eq. (\ref{eq:a1}), 
we can obtain the following relation
\begin{equation}
  \label{eq:a13}
  \frac{B_s}{\delta} = \frac{B_{in}}{L} \frac{\rho_s \gamma_s}{\rho_{in} \gamma_{in}}
  .
\end{equation}
Substitute this relation into Eq. (\ref{eq:a12}), 
we obtain 
\begin{equation}
  \label{eq:a14}
  2 \frac{B_{in}^2}{\gamma_{in}^2} \frac{\gamma_s^2 v_s^2}{L} \sim \frac{B_{in}^2}{L} \frac{\rho_s \gamma_s}{\rho_{in} \gamma_{in}} - \frac{p_o - p_N}{L}
  .
\end{equation}
This equation reduces to 
\begin{equation}
  \label{eq:a15}
 \gamma_s^2 v_s^2 \sim \frac{\rho_s \gamma_s \gamma_{in}}{2 \rho_{in}} - \frac{\gamma_{in}^2}{B_{in}^2} (p_o - p_N)
  .
\end{equation}
From the mass conservation equation Eq. (\ref{eq:a5}) and Eq. (\ref{eq:a4}), 
we can obtain the following relation: 
\begin{equation}
  \label{eq:a16}
  \frac{\rho_s \gamma_s}{\rho_{in}} \sim \frac{\gamma_{in} v_{in}}{v_s} \frac{L}{\delta} \sim \frac{\gamma_{in} v_{in}^2}{v_s} S_l
  .
\end{equation}
Using this equation, 
Eq. (\ref{eq:a15}) can be rewritten as follows: 
\begin{equation}
  \label{eq:a17}
% \gamma_s^2 v_s^2 \sim \frac{\gamma_{in}}{2} \frac{\gamma_{in} v_{in}^2}{v_s} S_l
  \gamma_s \sim \gamma_A \sqrt{1 + \frac{1}{2} \left[ 1 - \frac{p_o}{p_N} \right]} \equiv \gamma_A \alpha
  ,
\end{equation}
where $\gamma_A$ is the Lorentz factor of the Alfv\'en velocity in the upstream region 
and $\alpha$ is the effect of the pressure gradient. 
Note that when $p_0 > 3 p_N$, $\alpha$ becomes imaginary number. 
This is because in this case the reconnection outflow is prevented by the pressure gradient force 
and this means the break down of the assumption of the steady state. 
%This reduces to 
%\begin{equation}
%  \label{eq:a18}
% \gamma_{in} v_{in} \sim \sqrt{\frac{2}{S_l}} \gamma_s v_{s}^{3/2}
%  .
%\end{equation}
Using this equation and Eq. (\ref{eq:a11}), 
%we can obtain the following relations:
%\begin{eqnarray}
%  \label{eq:a19}
%  v_s &\sim& c_A
%  ,
%  \\
%  \label{eq:a20}
%  \gamma_{in} v_{in} &\sim& \sqrt{\frac{2}{S_l}} \sqrt{\sigma_{in} c_A}
%  \label{eq:a21}
%  .
%\end{eqnarray}
%Rewriting Eq. (\ref{eq:a21}), 
we obtain the following form of the reconnection rate: 
\begin{equation}
%  \gamma_{in} v_{in} / c_A \sim \sqrt{\frac{2}{S_L}} \sqrt{\sigma_{in}}
  \gamma_{in} v_{in} / c_A \sim \sqrt{\frac{2}{S_L}} \sqrt{\gamma_A c_A \alpha \sqrt{\gamma_A^2 \alpha^2 - 1}} 
  \label{eq:a22}
  , 
\end{equation}
where $S_L \equiv L c_A / \eta$ is the Lundquist number relating to the Alfv\'en velocity. 
When $p_o = p_N$, 
the above equation reduces to 
\begin{equation}
  \gamma_{in} v_{in} / c_A \sim \sqrt{\frac{2}{S_L}} \sqrt{\sigma_{in}}
  \label{eq:a23}
  , 
\end{equation}
which is equivalent to the relation obtained in \citep{2003ApJ...589..893L}. 
When $p_o = 0$, or the reconnection outflow is ejected into very cold region, 
Eq. (\ref{eq:a22}) means 
the reconnection rate is enhanced due to the pressure gradient force. 
Finally, 
when $p_o > p_N$, which is, for example, the case where plasmoids are existed at the edge of the current sheet, 
Eq. (\ref{eq:a22}) means 
the reconnection rate is reduced by the pressure gradient force. 


\begin{thebibliography}{59}
\expandafter\ifx\csname natexlab\endcsname\relax\def\natexlab#1{#1}\fi

\bibitem[{{Amano} \& {Kirk}(2013)}]{2013arXiv1303.2702A}
{Amano}, T., \& {Kirk}, J.~G. 2013, ArXiv e-prints

\bibitem[{{Barkov} \& {Baushev}(2011)}]{2011NewA...16...46B}
{Barkov}, M.~V., \& {Baushev}, A.~N. 2011, New A, 16, 46

\bibitem[{{B{\'a}rta} {et~al.}(2011{\natexlab{a}}){B{\'a}rta}, {B{\"u}chner},
  {Karlick{\'y}}, \& {Kotr{\v c}}}]{2011ApJ...730...47B}
{B{\'a}rta}, M., {B{\"u}chner}, J., {Karlick{\'y}}, M., \& {Kotr{\v c}}, P.
  2011{\natexlab{a}}, \apj, 730, 47

\bibitem[{{B{\'a}rta} {et~al.}(2011{\natexlab{b}}){B{\'a}rta}, {B{\"u}chner},
  {Karlick{\'y}}, \& {Sk{\'a}la}}]{2011ApJ...737...24B}
{B{\'a}rta}, M., {B{\"u}chner}, J., {Karlick{\'y}}, M., \& {Sk{\'a}la}, J.
  2011{\natexlab{b}}, \apj, 737, 24

\bibitem[{{Bessho} \& {Bhattacharjee}(2012)}]{2012ApJ...750..129B}
{Bessho}, N., \& {Bhattacharjee}, A. 2012, \apj, 750, 129

\bibitem[{{Bhattacharjee} {et~al.}(2009){Bhattacharjee}, {Huang}, {Yang}, \&
  {Rogers}}]{2009PhPl...16k2102B}
{Bhattacharjee}, A., {Huang}, Y.-M., {Yang}, H., \& {Rogers}, B. 2009, Physics
  of Plasmas, 16, 112102

\bibitem[{{Biskamp}(2000)}]{2000mrp..book.....B}
{Biskamp}, D. 2000, {Magnetic Reconnection in Plasmas}, ed. {Biskamp, D.}

\bibitem[{{Blackman} \& {Field}(1994)}]{1994PhRvL..72..494B}
{Blackman}, E.~G., \& {Field}, G.~B. 1994, Physical Review Letters, 72, 494

\bibitem[{{Dumbser} \& {Zanotti}(2009)}]{2009JCoPh.228.6991D}
{Dumbser}, M., \& {Zanotti}, O. 2009, Journal of Computational Physics, 228,
  6991

\bibitem[{{Evans} \& {Hawley}(1988)}]{1988ApJ...332..659E}
{Evans}, C.~R., \& {Hawley}, J.~F. 1988, \apj, 332, 659

\bibitem[{{Fermo} {et~al.}(2010){Fermo}, {Drake}, \&
  {Swisdak}}]{2010PhPl...17a0702F}
{Fermo}, R.~L., {Drake}, J.~F., \& {Swisdak}, M. 2010, Physics of Plasmas, 17,
  010702

\bibitem[{{Fermo} {et~al.}(2011){Fermo}, {Drake}, {Swisdak}, \&
  {Hwang}}]{2011JGRA..116.9226F}
{Fermo}, R.~L., {Drake}, J.~F., {Swisdak}, M., \& {Hwang}, K.-J. 2011, Journal
  of Geophysical Research (Space Physics), 116, 9226

\bibitem[{{Fujimoto}(2011)}]{2011PhPl...18k1206F}
{Fujimoto}, K. 2011, Physics of Plasmas, 18, 111206

\bibitem[{{Furth} {et~al.}(1963){Furth}, {Killeen}, \&
  {Rosenbluth}}]{1963PhFl....6..459F}
{Furth}, H.~P., {Killeen}, J., \& {Rosenbluth}, M.~N. 1963, Physics of Fluids,
  6, 459

\bibitem[{{Gardiner} \& {Stone}(2005)}]{2005JCoPh.205..509G}
{Gardiner}, T.~A., \& {Stone}, J.~M. 2005, Journal of Computational Physics,
  205, 509

\bibitem[{{Gardiner} \& {Stone}(2008)}]{2008JCoPh.227.4123G}
---. 2008, Journal of Computational Physics, 227, 4123

\bibitem[{{Giannios}(2013)}]{2013MNRAS.431..355G}
{Giannios}, D. 2013, \mnras, 431, 355

\bibitem[{{Goodman} \& {Uzdensky}(2008)}]{2008ApJ...688..555G}
{Goodman}, J., \& {Uzdensky}, D. 2008, \apj, 688, 555

\bibitem[{{Hoh}(1966)}]{1966PhFl....9..277H}
{Hoh}, F.~C. 1966, Physics of Fluids, 9, 277

\bibitem[{{Huang} \& {Bhattacharjee}(2012)}]{2012PhRvL.109z5002H}
{Huang}, Y.-M., \& {Bhattacharjee}, A. 2012, Physical Review Letters, 109,
  265002

\bibitem[{{Huang} \& {Bhattacharjee}(2013)}]{2013arXiv1301.0331H}
---. 2013, ArXiv e-prints

\bibitem[{{Inoue}(2012)}]{2012ApJ...760...43I}
{Inoue}, T. 2012, \apj, 760, 43

\bibitem[{{Kennel} \& {Coroniti}(1984{\natexlab{a}})}]{1984ApJ...283..694K}
{Kennel}, C.~F., \& {Coroniti}, F.~V. 1984{\natexlab{a}}, \apj, 283, 694

\bibitem[{{Kennel} \& {Coroniti}(1984{\natexlab{b}})}]{1984ApJ...283..710K}
---. 1984{\natexlab{b}}, \apj, 283, 710

\bibitem[{{Kirk} \& {Skj{\ae}raasen}(2003)}]{2003ApJ...591..366K}
{Kirk}, J.~G., \& {Skj{\ae}raasen}, O. 2003, \apj, 591, 366

\bibitem[{{Komissarov} {et~al.}(2007{\natexlab{a}}){Komissarov}, {Barkov}, \&
  {Lyutikov}}]{2007MNRAS.374..415K}
{Komissarov}, S.~S., {Barkov}, M., \& {Lyutikov}, M. 2007{\natexlab{a}},
  \mnras, 374, 415

\bibitem[{{Komissarov} {et~al.}(2007{\natexlab{b}}){Komissarov}, {Barkov},
  {Vlahakis}, \& {K{\"o}nigl}}]{2007MNRAS.380...51K}
{Komissarov}, S.~S., {Barkov}, M.~V., {Vlahakis}, N., \& {K{\"o}nigl}, A.
  2007{\natexlab{b}}, \mnras, 380, 51

\bibitem[{{Kowal} {et~al.}(2009){Kowal}, {Lazarian}, {Vishniac}, \&
  {Otmianowska-Mazur}}]{2009ApJ...700...63K}
{Kowal}, G., {Lazarian}, A., {Vishniac}, E.~T., \& {Otmianowska-Mazur}, K.
  2009, \apj, 700, 63

\bibitem[{{Lazarian} \& {Vishniac}(1999)}]{1999ApJ...517..700L}
{Lazarian}, A., \& {Vishniac}, E.~T. 1999, \apj, 517, 700

\bibitem[{{Loureiro} {et~al.}(2005){Loureiro}, {Cowley}, {Dorland}, {Haines},
  \& {Schekochihin}}]{2005PhRvL..95w5003L}
{Loureiro}, N.~F., {Cowley}, S.~C., {Dorland}, W.~D., {Haines}, M.~G., \&
  {Schekochihin}, A.~A. 2005, Physical Review Letters, 95, 235003

\bibitem[{{Loureiro} {et~al.}(2012){Loureiro}, {Samtaney}, {Schekochihin}, \&
  {Uzdensky}}]{2012PhPl...19d2303L}
{Loureiro}, N.~F., {Samtaney}, R., {Schekochihin}, A.~A., \& {Uzdensky}, D.~A.
  2012, Physics of Plasmas, 19, 042303

\bibitem[{{Loureiro} {et~al.}(2007){Loureiro}, {Schekochihin}, \&
  {Cowley}}]{2007PhPl...14j0703L}
{Loureiro}, N.~F., {Schekochihin}, A.~A., \& {Cowley}, S.~C. 2007, Physics of
  Plasmas, 14, 100703

\bibitem[{{Loureiro} {et~al.}(2013){Loureiro}, {Schekochihin}, \&
  {Uzdensky}}]{2013PhRvE..87a3102L}
{Loureiro}, N.~F., {Schekochihin}, A.~A., \& {Uzdensky}, D.~A. 2013, \pre, 87,
  013102

\bibitem[{{Lovelace} \& {Romanova}(2003)}]{2003ApJ...596L.159L}
{Lovelace}, R.~V.~E., \& {Romanova}, M.~M. 2003, \apjl, 596, L159

\bibitem[{{Low}(1973)}]{1973ApJ...181..209L}
{Low}, B.~C. 1973, \apj, 181, 209

\bibitem[{{Lyubarsky} \& {Kirk}(2001)}]{2001ApJ...547..437L}
{Lyubarsky}, Y., \& {Kirk}, J.~G. 2001, \apj, 547, 437

\bibitem[{{Lyubarsky}(2005)}]{2005MNRAS.358..113L}
{Lyubarsky}, Y.~E. 2005, \mnras, 358, 113

\bibitem[{{Lyutikov} \& {Blandford}(2003)}]{2003astro.ph.12347L}
{Lyutikov}, M., \& {Blandford}, R. 2003, ArXiv Astrophysics e-prints

\bibitem[{{Lyutikov} \& {Uzdensky}(2003)}]{2003ApJ...589..893L}
{Lyutikov}, M., \& {Uzdensky}, D. 2003, \apj, 589, 893

\bibitem[{{Mochol} \& {Kirk}(2013{\natexlab{a}})}]{2013arXiv1303.6434M}
{Mochol}, I., \& {Kirk}, J.~G. 2013{\natexlab{a}}, ArXiv e-prints

\bibitem[{{Mochol} \& {Kirk}(2013{\natexlab{b}})}]{2013arXiv1303.6781M}
---. 2013{\natexlab{b}}, ArXiv e-prints

\bibitem[{{Parker}(1957)}]{1957JGR....62..509P}
{Parker}, E.~N. 1957, \jgr, 62, 509

\bibitem[{{Parker}(1963)}]{1963ApJS....8..177P}
---. 1963, \apjs, 8, 177

\bibitem[{{Priest} \& {Forbes}(2000)}]{2000mare.book.....P}
{Priest}, E., \& {Forbes}, T. 2000, {Magnetic Reconnection}, ed. {Priest, E.~\&
  Forbes, T.}

\bibitem[{{Samtaney} {et~al.}(2009){Samtaney}, {Loureiro}, {Uzdensky},
  {Schekochihin}, \& {Cowley}}]{2009PhRvL.103j5004S}
{Samtaney}, R., {Loureiro}, N.~F., {Uzdensky}, D.~A., {Schekochihin}, A.~A., \&
  {Cowley}, S.~C. 2009, Physical Review Letters, 103, 105004

\bibitem[{{Shibata} \& {Tanuma}(2001)}]{2001EP&S...53..473S}
{Shibata}, K., \& {Tanuma}, S. 2001, Earth, Planets, and Space, 53, 473

\bibitem[{{Sweet}(1958)}]{1958IAUS....6..123S}
{Sweet}, P.~A. 1958, in IAU Symposium, Vol.~6, Electromagnetic Phenomena in
  Cosmical Physics, ed. B.~{Lehnert}, 123

\bibitem[{{Takahashi} {et~al.}(2011){Takahashi}, {Kudoh}, {Masada}, \&
  {Matsumoto}}]{2011ApJ...739L..53T}
{Takahashi}, H.~R., {Kudoh}, T., {Masada}, Y., \& {Matsumoto}, J. 2011, \apjl,
  739, L53

\bibitem[{{Takamoto} \& {Inoue}(2011)}]{2011ApJ...735..113T}
{Takamoto}, M., \& {Inoue}, T. 2011, \apj, 735, 113

\bibitem[{{Takamoto} {et~al.}(2012){Takamoto}, {Inoue}, \&
  {Inutsuka}}]{2012ApJ...755...76T}
{Takamoto}, M., {Inoue}, T., \& {Inutsuka}, S.-i. 2012, \apj, 755, 76

\bibitem[{{Ugai} \& {Zheng}(2005)}]{2005PhPl...12i2312U}
{Ugai}, M., \& {Zheng}, L. 2005, Physics of Plasmas, 12, 092312

\bibitem[{{Uzdensky} {et~al.}(2010){Uzdensky}, {Loureiro}, \&
  {Schekochihin}}]{2010PhRvL.105w5002U}
{Uzdensky}, D.~A., {Loureiro}, N.~F., \& {Schekochihin}, A.~A. 2010, Physical
  Review Letters, 105, 235002

\bibitem[{{Uzdensky} \& {Spitkovsky}(2012)}]{2012arXiv1210.3346U}
{Uzdensky}, D.~A., \& {Spitkovsky}, A. 2012, ArXiv e-prints

\bibitem[{{Zanotti} \& {Dumbser}(2011)}]{2011MNRAS.418.1004Z}
{Zanotti}, O., \& {Dumbser}, M. 2011, \mnras, 418, 1004

\bibitem[{{Zenitani} {et~al.}(2009{\natexlab{a}}){Zenitani}, {Hesse}, \&
  {Klimas}}]{2009ApJ...705..907Z}
{Zenitani}, S., {Hesse}, M., \& {Klimas}, A. 2009{\natexlab{a}}, \apj, 705, 907

\bibitem[{{Zenitani} {et~al.}(2009{\natexlab{b}}){Zenitani}, {Hesse}, \&
  {Klimas}}]{2009ApJ...696.1385Z}
---. 2009{\natexlab{b}}, \apj, 696, 1385

\bibitem[{{Zenitani} {et~al.}(2010){Zenitani}, {Hesse}, \&
  {Klimas}}]{2010ApJ...716L.214Z}
---. 2010, \apjl, 716, L214

\bibitem[{{Zenitani} \& {Miyoshi}(2011)}]{2011PhPl...18b2105Z}
{Zenitani}, S., \& {Miyoshi}, T. 2011, Physics of Plasmas, 18, 022105

\bibitem[{{Zhang} \& {Yan}(2011)}]{2011ApJ...726...90Z}
{Zhang}, B., \& {Yan}, H. 2011, \apj, 726, 90

\end{thebibliography}
\end{document}